\documentclass[11pt,a4paper]{article}
\usepackage[a4paper,margin=1.9cm]{geometry}
\usepackage{authblk}

\usepackage{appendix}
\usepackage{cite}
\usepackage{verbatim}
\usepackage{physics}
\usepackage{bbold}
\usepackage{xcolor}
\usepackage{ragged2e}
\usepackage[percent]{overpic}
\usepackage{orcidlink}
\usepackage{subcaption}
\usepackage{dcolumn}
\usepackage{amsmath,amssymb,amsfonts,nccmath}
\usepackage{bm}

\usepackage{hyperref}

\hypersetup{colorlinks=true, linkcolor=blue, filecolor=blue, citecolor = blue, urlcolor=blue}
\usepackage{chngcntr}
\usepackage{lipsum}
\usepackage{mathtools}
\usepackage{lmodern}

\newcommand{\floor}[1]{\lfloor #1 \rfloor}

\newcommand{\sigmap}{\hat{\sigma}^{+}}
\newcommand{\sigmaz}{\hat{\sigma}^{z}}
\newcommand{\sigmax}{\hat{\sigma}^{x}}

\newcommand{\A}{\hat{a}}
\newcommand{\Ac}{\hat{a}^{\dag}}
\newcommand{\B}{\hat{b}}
\newcommand{\Bc}{\hat{b}^{\dag}}

\newcommand{\Msin}[1]{\sin(#1)}
\newcommand{\Mcos}[1]{\cos(#1)}
\newcommand{\Mcosh}[1]{\cosh(#1)}
\newcommand{\Msinh}[1]{\sinh(#1)}
\newcommand{\Mtan}[1]{\tan(#1)}

\newcommand{\Msinq}[1]{\sin^2\left(#1\right)}

\newcommand{\df}{\partial_{\phi}}
\newcommand{\epsn}{\varepsilon_n(\phi)}
\newcommand{\psidw}{\downarrow}
\newcommand{\psiup}{\uparrow}
\newcommand{\psiO}{\ket{\psi(0)}}
\newcommand{\psit}{\ket{\psi(t)}}
\newcommand{\omegar}{\hbar \omega_{\rm r} }

\newcommand{\deltaO}{\Delta_{0}}

\newcommand{\kf}{k_{\rm F}}
\newcommand{\hvf}{\hbar v_{\rm F}}

\definecolor{dgr}{rgb}{0.0, 0.45, 0.0}

\begin{document}










\title{Quantum Batteries in two-dimensional material-based Josephson Junctions}

\author[1,2]{V. Varrica~\orcidlink{0009-0002-2424-0044}}
\author[3,4]{G. Gemme~\orcidlink{0009-0003-4942-6643}}
\author[1,2]{F.M.D. Pellegrino~\orcidlink{0000-0001-5425-1292}}
\author[1,2]{E. Paladino~\orcidlink{0000-0002-9929-3768}}
\author[3,4]{M. Sassetti~\orcidlink{0000-0002-1974-8209}}
\author[3,4]{D. Ferraro~\orcidlink{0000-0002-4435-1326}}

\affil[1]{{\small Dipartimento di Fisica e Astronomia "Ettore Majorana", Universit\`a di Catania, Via S. Sofia 64, I-95123 Catania,~Italy}}
\affil[2]{{\small INFN, Sez.~Catania, I-95123 Catania,~Italy}}
\affil[3]{{\small Dipartimento di Fisica, Universit\`a di Genova, Via Dodecaneso 33, I-16146 Genova,~Italy}}
\affil[4]{{\small CNR-SPIN, Via Dodecaneso 33, I-16146 Genova,~Italy}}

\date{}
\maketitle

\begin{abstract}
We investigate the solid-state implementation of a Dicke-like quantum battery consisting of a two-dimensional material-based Josephson junction inductively coupled to a resonator, using graphene as a representative example.
In this configuration, Andreev bound states naturally act as non-interacting, energetically non-degenerate two-level systems, and the setup allows for both single-photon and two-photon resonant processes.
The coupling between the LC‑circuit flux and the supercurrent through the junction gives rise to peculiar longitudinal interaction terms that have no counterpart in the conventional Dicke model. We demonstrate that these additional couplings can enhance energy storage for a proper range of parameters.
The proposed architecture also enables an alternative, but comparable, charging protocol that relies on tuning the superconducting phase difference across the junction.
\end{abstract}

\section{Introduction}\label{sec:intro}
Quantum batteries (QBs) are miniaturized devices able to efficiently store, transfer, and release energy on-demand by exploiting purely quantum effects~\cite{quach23, Campaioli24, Camposeo25, Ferraro26, Wang26}. They are progressively assuming a prominent role in the domain of quantum technologies~\cite{ezratty2025understandingquantumtechnologies2025} and in the emerging field of quantum energy~\cite{Auffeves22}. Their main aim is to provide an energy supply to quantum devices and quantum sensors supporting their increasing complexity~\cite{Chiribella21, Kurman25}.

The first theoretical proposals of QBs were based on a collection of independent two-level systems (TLSs)~\cite{Alicki13, Binder15}. Starting from these seminal works the field rapidly evolved, addressing a plethora of diverse platforms such as quantum spin chains~\cite{Le18, Rossini19, Barra22, Grazi24, Catalano24, Grazi25, Lu25, Ho26}, nuclear spins~\cite{Joshi22, Cruz22}, collisional models~\cite{Seah21, Shaghaghi22, Shaghaghi23, Morrone23, Massa25, Elyasi25}, superconducting circuits~\cite{Hu22, Gemme24,razzoli2025cyclic, Li25} and quantum harmonic oscillators~\cite{Hovhannisyan20, Cavaliere25, Cavaliere26}. Within this extremely varied context, probably the most successful scheme proposed so far is based on the matter-radiation interaction described by the Dicke model~\cite{dicke1954, kirton2018}. Here, a collection of independent TLSs composing the QB is embedded in a resonant cavity that acts as a charger~\cite{ferraro2018high, crescente2020ultrafast,Ferraro19, Carrasco22, gemme2023off, Yang24, Sharma25, Hadipour2026}. The charging process then consists in destroying the photons and promoting a large majority of the TLSs from the ground to the excited state. These Dicke-QBs show a collective quantum advantage in the charging power~\cite{Andolina19, Julia20}, which has been experimentally observed in molecular-based setups~\cite{quach2022micro}. The latest developments in this direction have addressed two fundamental aspects with the aim of improving the performance of these devices at the level of charging and storage respectively: the optimization of charging protocols using reinforcement learning techniques~\cite{Rodriguez23, Erdman24, Sun25} and the exploitation of the features of the energy spectrum of the molecules composing the QBs~\cite{Tibben25, Hymas26}.

Due to the important achievements realized in the domain of circuit quantum electrodynamics (cQED)~\cite{Xiang13, Krantz19}, where unconventional regimes of matter-radiation interaction can be addressed using superconducting circuits coupled with resonators acting as LC circuits, it seems natural to investigate possible implementations of Dicke-QBs also in these solid-state based platforms. In this direction, the present paper aims to explore a two-dimensional material-based Josephson junction inductively coupled with an LC circuit.

Although the considered formalism applies to any planar hybrid JJ, the graphene Josephson junction (GJJ) has been selected as a representative and experimentally well-established two-dimensional platform which has been successfully integrated
with microwave resonators and superconducting circuits, including
flux-tunable transmons and microwave cavities~
\cite{schmidt2018,kroll2018,wang2019}.
Within the short junction limit, single- and two-photon Dicke-QB physics naturally emerge with Andreev bound states (ABSs) playing the role of non-interacting and energetically non-degenerate TLSs. Due to the presence of interaction between the LC-circuit flux and the supercurrent flowing through the junction, longitudinal couplings with no correspondence in the conventional Dicke model emerge. They lead to an enhancement of the energy storage associated with two-photon processes while influencing the dynamics of single-photon resonant processes. More generally, the developed formalism is based on the presence of ABSs in a short hybrid JJ, and therefore
it can be applied to other two-dimensional (2D) hybrid systems, such as semiconductor 2D electron gases or van der Waals heterostructures~\cite{kjaergaard2017hetero,marcus2023planar,strickland2024}, provided that the junction
operates in the short-junction regime. At the current stage, representative decoherence benchmarks for ABSs have been reported in several platforms. For instance, in superconducting atomic contacts based on InAs/Al, single-shot control of an Andreev pair qubit yields excited-state lifetimes $T_1\approx 1-10~\mu\mathrm{s}$ and coherence time $T_2 \approx 10-100~\mathrm{ns}$~\cite{janvier2015}.
Whereas, in gate- and flux-tunable InAs nanowire Josephson junctions operated as Andreev spin qubits, microwave measurements yield energy-relaxation times $T_1 \approx 10~\mu\mathrm{s}$ and coherence times in the range $T_{2} \approx 1-10~\mathrm{ns}$~\cite{hays2018,hays2020,bargebos2022,pitavidal2023}. Moreover, estimation of characteristic timescales for ABSs defined in a planar JJ based on InAs/Al heterostructures, hosting a 2D electron gas, reveals $T_1 \approx 600~\rm{ns}$ and  $T_2\approx 4.5~\rm{ns}$~\cite{hinderling2023}.  In planar Ge Josephson junctions defined in proximitized 2D hole gases, microwave spectroscopy, combined with real-time monitoring of a dispersively coupled resonator, reveals time scales which characterize ABSs parity stability of the order of several seconds and dephasing time $T_2\simeq 1~\mathrm{ns}$~\cite{hinderling2024}. Taken together, these results indicate that, depending on the specific hybrid platform and ABSs degrees of freedom adopted, ABS-based qubits currently achieve energy-relaxation and parity lifetimes up to tens of microseconds, while coherence times typically lie in the nanosecond range.

In the limit of a short junction, where the physical length of the junction $L$ is much smaller than the superconducting coherence length, $\xi \sim\hvf/\Delta_0$, in the JJ, the flow of a supercurrent is microscopically sustained by pairs of ABSs, spatially confined in the normal phase region, which arise as a result of Andreev reflections occurring at the superconductor-semiconductor interfaces~\cite{beenakker1991}.
In this work, the limitation of the short-junction regime is crucial because, once one enters the intermediate-junction regime, {\it i.e.} $L\lesssim \xi$, the subgap Andreev spectrum no longer consists of independent TLSs but instead forms multiplets. As a result, the effective low-energy theory undergoes a qualitative change and requires a more elaborate description.

This device allows for the implementation of a charging protocol based on the modulation of the superconducting phase difference across the junction. This makes our proposal particularly interesting in view of experimental realizations.

\section{Model}\label{sec:model}
\begin{figure}[t]
	\centering
    \includegraphics[width=0.4\textwidth]{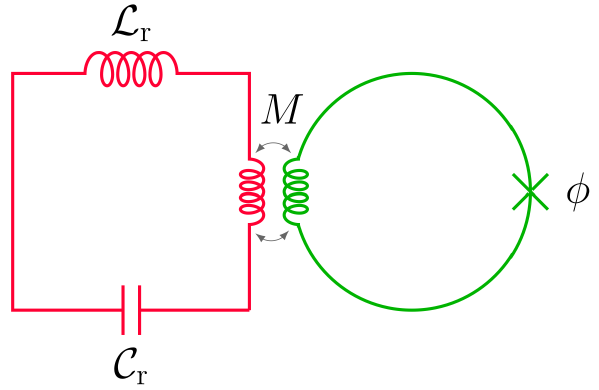}
	\caption{Scheme of a resonant circuit (red), described as a lumped-element LC circuit with an inductance $\mathcal{L}_{\rm r}$ and a capacitance $\mathcal{C}_{\rm r}$, playing role of a quantum charger. This LC circuit is coupled, through a mutual inductance $M$, with a ring containing a single short JJ (green), which represents the QB. Here, $\phi$ is the superconducting phase difference between the two superconducting leads.}
    \label{fig:GJJ_LC_circuit_schematic}
\end{figure}

We consider a device composed of a superconducting ring interrupted by a short superconductor-semiconductor Josephson Junction (JJ), which interacts via a mutual inductance $M$ with a superconducting resonator, as shown in Fig.~\ref{fig:GJJ_LC_circuit_schematic}.
In this cQED setup, the superconducting ring hosting the JJ (green) acts as the QB, while the superconducting resonator circuit (red) plays the role of the charger. Within this description, we are considering a charger-mediated protocol, in which the two quantum subsystems exchange energy through their mutual coupling, enabling the transfer of excitations between the charger and the QB~\cite{ferraro2018high, gemme2023off, quach2022micro}.

In a JJ in the short junction regime, ABSs are subgap states, namely eigenstates with energies symmetrically arranged around the Fermi level, whose level spacing is smaller than the induced superconducting energy gap $2\Delta_0$~\cite{zazunov2005}.
In realistic hybrid planar JJs, the finite transparency and quality of the semiconductor--superconductor interface results in an induced superconducting gap that is generally smaller than the bulk superconducting gap, with typical values of the order of $\Delta_0 \approx 0.1~\mathrm{meV}$ (corresponding to $\Delta_0/h \approx 20\,\mathrm{GHz}$)~\cite{kjaergaard2016,park2018,kjaergaard2017hetero}. The transport through JJ is also characterized by the total number of conduction channels $N$, which depends both on the microscopic properties of the semiconductor component and the geometry of the junction~\cite{beenakker1992}.
In the limit of a short junction, a single pair of ABSs is expected per conduction channel~\cite{janvier2015} with the eigenenergies $\pm\epsn$ with
\begin{equation}
	\epsn =\deltaO\sqrt{1-\tau_n\Msinq{\phi/2}}~,
	\label{eq:ABSs_energy}
\end{equation}
where $\pm$ denotes if the ABS energy is above ($+$) or below ($-$) the Fermi level, $\tau_n$ is the transmission probability of the $n$--th conduction channel, and $\phi$ is the superconducting phase difference between the two superconducting leads of the junction. As a result, at zero temperature, each populated ABS contributes to the current-phase relation (CPR) of the junction~\cite{bretheau2013}, which is given by
\begin{equation}
	I(\phi) = - \frac{2e}{\hbar}\sum_{n=1}^{N}\df\epsn~.
	\label{eq:CPR_hybrid_JJ}
\end{equation}

Within this description, the system of ABSs is represented as a collection of $N$ non-interacting TLSs, characterized by different energy splittings $2\epsn$. Specifically, for a given value of $\phi$, by projecting the matter Hamiltonian in the low-energy subspace spanned by the ABSs~\cite{pellegrino2022effect}, we can model the Hamiltonian of our QB system as
\begin{equation}
    \hat{H}_{\rm B} = \sum_{n=1}^{N}\epsn\,\sigmaz_{n}~,
    \label{eq:QB_hamiltonian}
\end{equation}
where $\hat{\sigma}^{\beta}_{n}$ ($\beta=x,y,z$) are the Pauli matrices associated with the $2\times2$ subspace spanned by the $n$--th pair of ABSs.
In line with this, we define two more relevant eigenstates of the QB Hamiltonian in Eq.~\eqref{eq:QB_hamiltonian}, which are expressed as product states
\begin{subequations}
    \begin{align}
    \ket{\psidw} &=  \mathop{\otimes}\limits_{n=1}^{N}\ket{\downarrow}_{n}~,
    \label{eq:QB_empty_full_states}
    \\ \ket{\psiup} &=  \mathop{\otimes}\limits_{n=1}^{N}\ket{\uparrow}_{n}~,
    \end{align}
\end{subequations}
where the first one is the ground state and the second one is the maximally excited eigenstate.

The auxiliary system, which plays the role of the charger, is a superconducting resonator (red) shown in Fig.~\ref{fig:GJJ_LC_circuit_schematic}. We model it as a lumped-element LC circuit with an inductance $\mathcal{L}_{\rm r}$ and a capacitance $\mathcal{C}_{\rm r}$.  Following the conventional quantization procedure~\cite{devoret2017}, the Hamiltonian of the LC circuit reads
\begin{equation}
    \hat{H}_{\rm r} = \omegar\left(\Ac\A +\frac{1}{2}\right)~,
    \label{eq:LC_hamiltonian}
\end{equation}
where $\A$ ($\Ac$) annihilates (creates) a photon of energy $\omegar$, with $\omega_{r}=1/\sqrt{\mathcal{L}_{\rm r}\,\mathcal{C}_{\rm r}}$. As outlined in previous works~\cite{park2020, hays2020, Varrica2026, metzger2021, elfeky2025}, the interaction, which arises from the inductive scheme shown in Fig.~\ref{fig:GJJ_LC_circuit_schematic}, can be expressed in terms of a dimensionless coupling constant $g = M\Phi_{\rm zpf}/(\mathcal{L}_{\rm r}\phi_0)$, where $\Phi_{\rm{zpf}} = \sqrt{\hbar/(2{\mathcal C}_{\rm r} \omega_{\rm r})}$ represents the zero-point flux fluctuations and $\phi_0 = \hbar/(2e)$ denotes the reduced flux quantum.
Assuming near-ideal mutual inductance with a ratio $M/{\cal L}_{\rm r}\lesssim 1$, and taking the inductance to be ${\cal L}_{\rm r}\sim 10~\text{nH}$ and the capacitance to be ${\cal C}_{\rm r}\sim 1~\text{pF}$, the coupling constant can reach values up to $g\lesssim 0.3$, which are comparable to those chosen in the following analysis.
Within this picture, the total (charger+QB) Hamiltonian is obtained as a second-order expansion with respect to the coupling \cite{metzger2021,Varrica2026} and reads
\begin{equation}
    \begin{aligned}
    \hat{H} &=\hat{H}_{\rm r}+ \hat{H}_{\rm B}+ g (\A + \Ac)\sum_{n=1}^{N}\vb{P}_{n}\cdot \hat{\boldsymbol{\sigma}}_{n}+\frac{g^2}{2} \left(\A +\Ac\right)^2\sum_{n=1}^{N}\vb{D}_{n}\cdot \hat{\boldsymbol{\sigma}}_{n}~,
    \end{aligned}
    \label{eq:compact_Hamiltonian_tot}
\end{equation}
where we introduced the short notations $\hat{\boldsymbol{\sigma}}_{n}=\left(\sigmaz_{n},\,\sigmax_{n}\right)^{\rm T}$ and
\begin{equation}
    \begin{aligned}
        \vb{P}_{n}&= \left(P_n^{z}(\phi),\,P_n^{x}(\phi)\right)^{T} =\df\epsn\left(1,\,-\sqrt{1-\tau_n}\Mtan{\frac{\phi}{2}}\right)^{\rm T}~,\\
        \vb{D}_{n} &= \left(D_n^{z}(\phi),\,D_n^{x}(\phi)\right)^{T}=\df\epsn\left(\frac{\tau_n +(2-\tau_n) \Mcos{\phi}}{2\Msin{\phi}},\,-\sqrt{1-\tau_n}\right)^{T}~,
    \end{aligned}
    \label{eq:ABSs_paramg_diamg_terms}
\end{equation}
as vectors which encode the single-photon and two-photon coupling terms, respectively~\cite{Felicetti15, Felicetti18}. Indeed, the third and fourth terms in Eq.~\eqref{eq:compact_Hamiltonian_tot} are analogous to the
paramagnetic and diamagnetic contributions which typically emerge in the light-matter coupling Hamiltonian~\cite{Nataf10, andolina2019}. Moreover, both contributions are proportional to the derivative of the $n$-th ABS energy $\epsn$ with respect to the superconducting phase difference $\phi$, introduced in Eq.~\eqref{eq:ABSs_energy}, namely
\begin{equation}
\df\epsn=-\frac{\Delta_{0}}{4}\frac{\tau_{n} \sin{\phi}}{\sqrt{1-\tau_n\Msinq{\phi/2}}}.
\end{equation}

We emphasize that the transmission probability $\tau_n$ enters the description of the QB at two distinct levels: it fixes the bare energy
splitting $2\varepsilon_n(\phi)$ of each ABS pair through Eq.~\eqref{eq:ABSs_energy}, and it
simultaneously determines the single- and two-photon coupling strengths $g\vb{P}_{n}$ and $g^2\vb{D}_{n}/2$, which affect how efficiently each ABS pair exchanges energy with the resonator mode. In general, the distribution of $\tau_n$ depends both on the microscopic properties of the semiconductor component, such as its doping level, and by the junction geometry \cite{goffman2017,nichele2020relating}, including details like the terminations~\cite{bonasera_arxiv_2026}. Consequently, a realistic platform based on a planar hybrid JJ exhibits non-uniform distributions of both energy splittings and effective coupling strengths. This is radically different from the single, homogeneous TLS ensemble assumed in a conventional Dicke
model.
\begin{figure*}[t]
  \centering
  \begin{subfigure}[c]{0.40\textwidth}
    \caption{}
    \includegraphics[width=\textwidth]{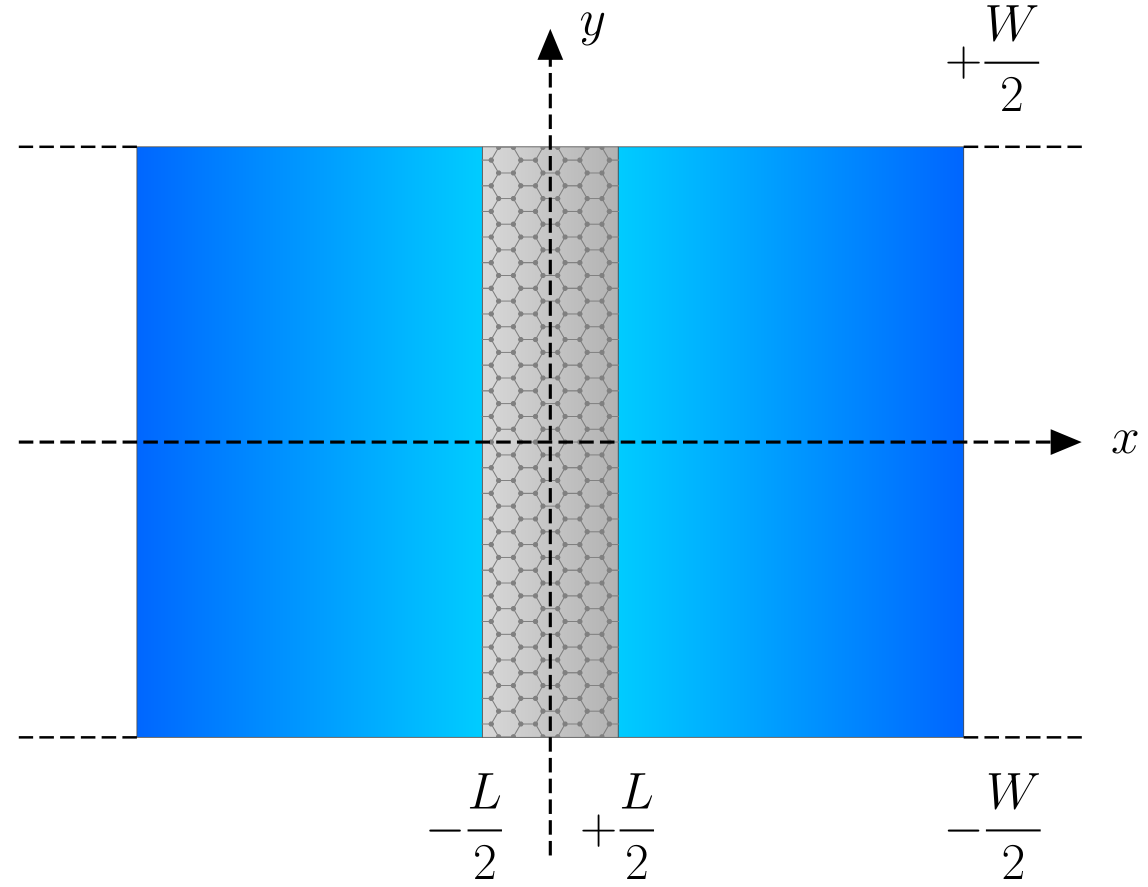}
    \label{fig:GJJ_plan_lattice}
  \end{subfigure}
  \begin{subfigure}[c]{0.49\textwidth}
    \caption{}
    \includegraphics[width=\textwidth]{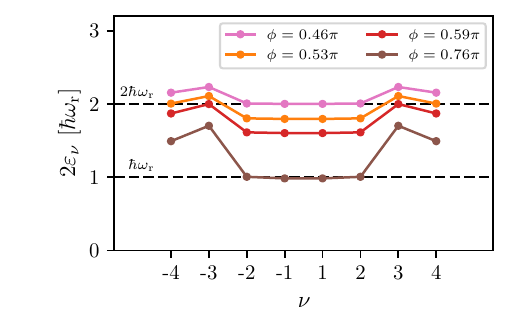}
    \label{fig:QB_energies}
  \end{subfigure}
  \caption{\subref{fig:GJJ_plan_lattice} Top view of a Josephson junction formed by graphene covered by two superconducting leads (blue). The uncovered grey region represents the graphene stripe in the normal phase. In this picture, $L$ represents the junction channel length along the $x$-direction and $W$ is the width of the device along the $y$-direction.
  \subref{fig:QB_energies} The energy splitting of each ABSs pair evaluated at $\phi/\pi=0.46$ (pink), $\phi/\pi=0.53$ (orange), $\phi/\pi=0.59$ (red), $\phi/\pi=0.76$ (brown). Horizontal dashed lines indicate the energies $\omegar$ and $2\omegar$, which make it possible to identify the units tuned to single-photon or two-photon resonance conditions, respectively.
  Here, the index $\nu=\zeta m$ is expressed using negative (positive) integers corresponding to the valley index $\zeta=-$ ($\zeta=+$), where $m$ labels the associated wavevector $q_m$ as defined in Eq.~\eqref{eq:q_wavenumber}.
  Other parameters are $\omegar=0.75\deltaO$, $\mu_0 = 6.5\hvf/L$ and $W = 2 L$.
  }
  \label{fig:GJJ_Energies}
\end{figure*}

Concerning the charging protocol, we follow what is usually proposed in the framework of Dicke QBs~\cite{ferraro2018high, andolina2018, quach2022micro, Kurman25}, which consists of a sudden quench of the light-matter coupling. An alternative protocol implementable in the present setup will be discussed at the end of the paper. Consequently, we assume a time modulation of the inductive coupling in Eq.~\eqref{eq:compact_Hamiltonian_tot} in the form
\begin{equation}
	g \to \bar{g}(t) =  g\Theta(t)\Theta(t_{\rm c}-t)~,
	\label{eq:g_protocol}
\end{equation}
where $\Theta(t)$ is the Heaviside step function and $t_{\rm c}$ is a controllable charging time.

In this work, we focus on the energy stored in the QB at a given time $t$ as a figure of merit to characterize the performance of the device. It is defined as~\cite{andolina2018}
\begin{equation}
    \begin{aligned}
    E^{(g)}_{\rm B}(t) = \expval{\hat{H}_{\rm B}}{\psi (t)} -  \expval{\hat{H}_{\rm B}}{\psi(0)},
    \end{aligned}
    \label{eq:QB_stored_energy}
\end{equation}
where $\psiO$ denotes the initial state of the global system (see below), while $\psit$ represents the state at time $t\leq t_{\rm c}$, which evolved according to the unitary generated by the Hamiltonian in Eq.~\eqref{eq:compact_Hamiltonian_tot}.
Notice that the present approach assumes unitary dynamics and is therefore valid whenever relaxation and dephasing can be neglected during the charging process. Beyond this regime, an open-system description becomes necessary~\cite{carrega_njp_2020}. However, recent studies have shown that decoherence does not necessarily hinder the performance of QBs and may even enhance the charging under suitable conditions~\cite{Cavaliere25,Cavaliere26,shastri_npj_2025}. Although these results cannot be directly extended to the present approach, they suggest that dissipation and dephasing could not represent a fundamental limitation on the QB performances.

As stated above, in contrast to the conventional Dicke model~\cite{dicke1954,kirton2018} which involves a set of identical TLSs, the QB considered here is composed of TLSs with a distribution of energy splittings and coupling strengths \cite{strter2012,diniz2011,tsyplyatyev2009}.
This has a deep impact on the symmetries of the system. More specifically, if we consider the collective pseudospin operators $\hat{S}_{\beta} = (\hbar/2)\sum_{n=1}^{N}\hat{\sigma}^{\beta}_{n}$, in our case we have that
\begin{equation}
    \comm*{\hat{S}^2}{\hat{H}(t)}\neq 0~,
    \label{eq:QB_comm_pseudosp_operators}
\end{equation}
where $\hat{S}^2 = \hat{S}^2_{x}+\hat{S}^2_{y}+\hat{S}^2_{z}$.
Therefore, we cannot take advantage of the block decomposition of the Dicke Hamiltonian based on the conservation of $S^2$ and the consequent reduction in computational cost. This leads us to investigate the behaviour of the QB, containing $N$ ABSs, employing a fully numerical procedure which addresses the complete Hilbert space of the ABSs (matter) sector characterized by a dimension $d_{\rm m}=2^N$. In addition to this, although a rigorous treatment of the radiation sector would involve an infinite Hilbert space, we employ a truncated representation of dimension $d_{\rm ph}$ that accommodates a maximum of $4N$ excitations. We have set this as the dimension threshold after a comparison with the results obtained in the case with $d_{\rm ph}+1$, which confirmed that this dimension is large enough to guarantee an accurate convergence of the numerical results~\cite{ferraro2018high, gemme2023off}.


\section{Results}\label{sec:results}
In this work, we consider a short planar graphene JJ (GJJ) characterized by a finite width $W$, as shown in Fig.~\ref{fig:GJJ_plan_lattice}. The electronic properties of this device can be tuned by changing the Fermi level $\mu_0$.
By exploiting the device spatial symmetry, it is natural to use the transverse momentum component to label the system eigenstates when solving the eigenvalue problem. Due to the finite width $W$, it is also necessary to consider boundary conditions. Specifically, following Ref.~\cite{titovPRB2006}, we apply the so-called infinite-mass boundary conditions for the $y=\{-W/2,\,+W/2\}$ transverse boundaries. As a result, the transverse wavevector $k$ becomes quantized as
\begin{equation}
    q_m = \dfrac{\pi L}{W}\left(m-\frac{1}{2}\right)~,\quad m = 1, 2, \dots~,
    \label{eq:q_wavenumber}
\end{equation}
where we multiplied the transverse wavevector by $L$, in order to express $q_m$ in dimensionless form.

The short-junction regime, $L \ll \xi$, enables us to express the ABS eigenenergies in terms of the transmission probability within the normal-phase region~\cite{beenakker1991}.
In particular, for the GJJ under investigation, one has
\begin{equation}
\tau_m = \frac{(\kf^2- q_m^2)}{[\kf^{2} - q_m^2\cos^2(\sqrt{\kf^{2} -q_m^2}\,)]},
\label{eq:graphene_normal_phase_transmission}
\end{equation}
where $\kf = \abs{\mu_0}L/\hvf$ is the Fermi wavenumber in units of $1/L$ and $v_{\rm F}\sim c/300$ is the Fermi velocity in graphene.
From the dependence of $\tau_m$ on $\mu_0$, one can see that the level acts as a control parameter that allows one to modify the distribution of transmission probabilities. This is a characteristic feature of hybrid JJs, which constitute the central element of gatemons~\cite{kroll2018}.
Furthermore, the number of propagating modes, i.e., characterized by $q_m \leq \kf$, is given by $N_{\rm m} = \floor{\mu_{0}W/\pi\hvf + 1/2}$ ($\floor{\cdot}$ indicating the integer part). Notice that we have also taken into account the valley degree of freedom in graphene, $\zeta=\pm$, which
introduces a double degeneracy in our model. Consequently, $N\equiv2N_{\rm m}$ is the total number of conduction channels responsible for transport in the GJJ that coincide with the number of units composing the QB. According to this, comparing with the Hamiltonian in Eq.~\eqref{eq:compact_Hamiltonian_tot}, it is useful to replace the label $n$ with a composite index $\nu\equiv \zeta m $.

\begin{figure*}[t]
  \centering
  \begin{subfigure}[t]{0.485\textwidth}
    \caption{}
    \includegraphics[width=0.99\textwidth]{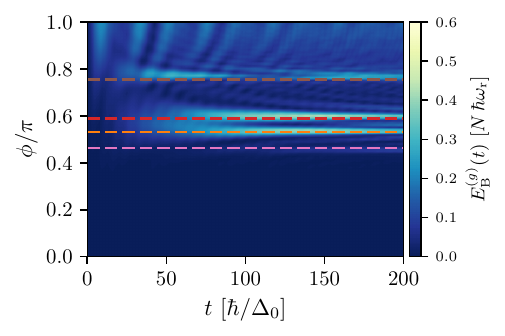}
    \label{fig:Energy_scan_g_0.100}
  \end{subfigure}
  \hfill
  \begin{subfigure}[t]{0.485\textwidth}
    \caption{}
    \includegraphics[width=0.99\textwidth]{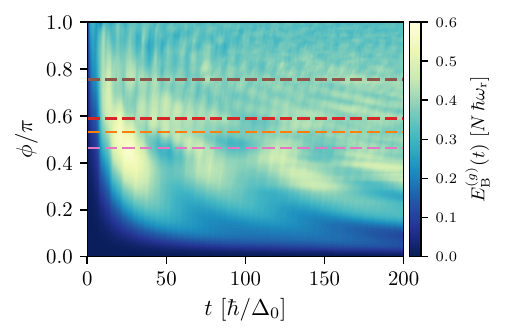}
    \label{fig:Energy_scan_g_0.300}
  \end{subfigure}
  \caption{Energy $E^{(g)}_{\mathrm{B}}$ stored in the QB system (in units of $N\omegar$) as a function of time and the  superconducting phase difference $\phi$, setting $g=0.1$ \subref{fig:Energy_scan_g_0.100} and $g=0.3$ \subref{fig:Energy_scan_g_0.300}.
 In both panels, using the same scheme of colors used in Fig.~\ref{fig:QB_energies}, the dashed lines indicate the values of $\phi$ such that there are at least two pairs of ABSs which enable single-photon ($\phi=0.76\pi$, brown), or two-photon resonant processes ($\phi=0.46 \pi$, pink), ($\phi=0.53 \pi$, orange), and ($\phi=0.59 \pi$ red).
  In all panels, the times are reported in units of $\hbar/\deltaO$. Other parameters are $\omegar=0.75\deltaO$, $\mu_0 = 6.5\hvf/L$ and $W = 2 L$.}
  \label{fig:Energy_scan_g}
\end{figure*}

In the following analysis, we will set the Fermi level and the geometric aspect ratio to the fixed values $\mu_0 = 6.5\,\hbar v_{\rm F}/L$ and $W/L = 2$, respectively, in line with typical experimental parameters~\cite{mayer2019,haller2022,turini2024}.
This experimentally feasible configuration yields a QB system with $N=8$, as shown in Fig.~\ref{fig:QB_energies}.
Once the Fermi level and the geometric aspect ratio are fixed, the set of transmission probabilities $\{\tau_n\}$ is completely determined by Eq.~\eqref{eq:graphene_normal_phase_transmission}.

Furthermore, the same Figure shows the dependence of the splitting energy of each ABSs pair on the superconducting phase difference across the junction, $\phi$. In particular, we set the bare resonator energy at $\hbar \omega_{\rm r}=0.75 \Delta_0$ and tune $\phi$ so that there are at least two pairs of ABSs fulfilling either the single-photon resonance condition $2 \varepsilon_n(\phi)=\hbar \omega_{\rm r}$ (shown in brown in Fig.~\ref{fig:QB_energies}) or the two-photon resonance condition $2 \varepsilon_n(\phi)=2\hbar \omega_{\rm r}$ (shown in pink, orange, and red in Fig.~\ref{fig:QB_energies}).
In what follows, we denote these cases as resonance conditions, whereby the two possible resonance processes occur selectively for at least two pairs of ABSs rather than for the entire ensemble.
\begin{figure*}[t]
  \centering
  \begin{subfigure}[t]{0.485\textwidth}
    \caption{}
    \includegraphics[width=0.99\textwidth]{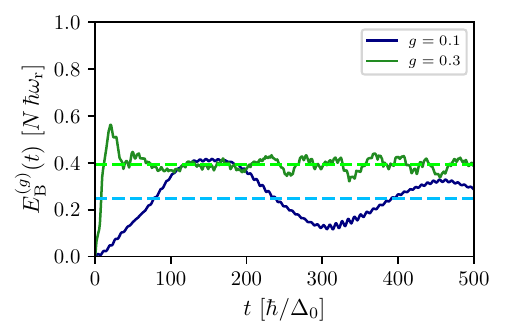}
    \label{fig:Energy_static_phi_0.53}
  \end{subfigure}
  \hfill
  \begin{subfigure}[t]{0.485\textwidth}
    \caption{}
    \includegraphics[width=0.99\textwidth]{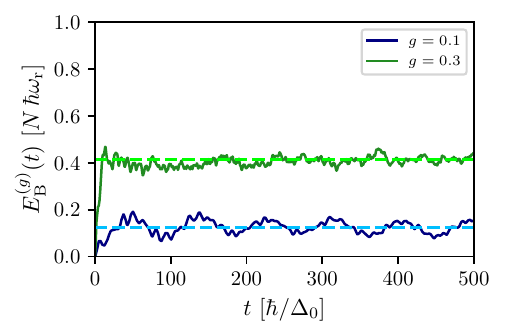}
    \label{fig:Energy_static_phi_0.76}
  \end{subfigure}

  \vspace{-2.0em}

  \begin{subfigure}[t]{0.485\textwidth}
    \caption{}
    \includegraphics[width=0.99\textwidth]{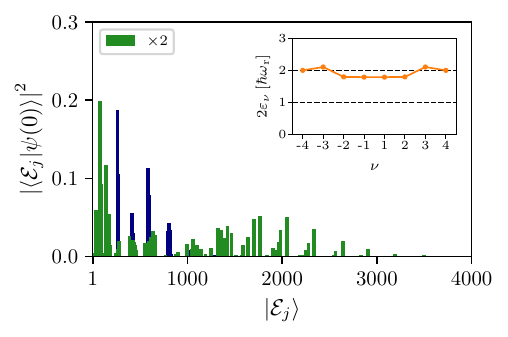}
    \label{fig:Abs_Overlap_phi_0.53}
  \end{subfigure}
  \hfill
  \begin{subfigure}[t]{0.485\textwidth}
    \caption{}
    \includegraphics[width=0.99\textwidth]{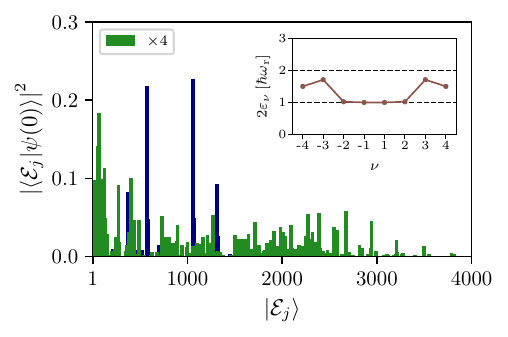}
    \label{fig:Abs_Overlap_phi_0.76}
  \end{subfigure}

  \caption{Long-time behaviour of $E^{(g)}_{\rm B}$ (solid lines) evaluated
  at $\phi/\pi=0.53$ \subref{fig:Energy_static_phi_0.53} and
  at $\phi/\pi=0.76$ \subref{fig:Energy_static_phi_0.76}. In both panels, time is expressed in units of $\hbar/\deltaO$, and the stationary values $\bar{E}^{(g)}_{\rm B}$ (dashed lines) are denoted in light green for $g=0.1$ and in cyan for $g=0.3$.
  \subref{fig:Abs_Overlap_phi_0.53}--\subref{fig:Abs_Overlap_phi_0.76} Bar plots illustrating the decomposition of the initial state $|\psi (0)\rangle$, over the relevant subset of the eigenbasis $\{\ket{\mathcal{E}_j}\}$, corresponding to the cases shown in \subref{fig:Energy_static_phi_0.53}--\subref{fig:Energy_static_phi_0.76}. To improve graphical readability, the green bars in  \subref{fig:Abs_Overlap_phi_0.53} and \subref{fig:Abs_Overlap_phi_0.76} are scaled by a factor $2$ and $4$, respectively. In the insets, the distribution of the ABSs energy splittings is shown (as in Fig.~\ref{fig:QB_energies}). In all panels, we compare two values of the coupling constant: $g=0.1$ (blue) and $g=0.3$ (green). Other parameters are $\omegar=0.75\deltaO$, $\mu_0 = 6.5\hvf/L$ and $W = 2 L$.
  }
  \label{fig:Diagonalization_tests}
\end{figure*}

We numerically investigate the time evolution of the stored energy $E^{(g)}_{\rm B}$, defined in Eq.~\eqref{eq:QB_stored_energy}, by considering different fixed values of the superconducting phase difference $\phi$. Here, we introduce the superscript ($g$) to indicate that the evolution is governed by the charging protocol defined in Eq.~\eqref{eq:g_protocol}, thus distinguishing it from the alternative charging protocol discussed below. In particular, the initial state is
\begin{equation}
    |\psi (0)\rangle = \ket{N}\otimes\ket{\psidw}~,
    \label{eq:initial_state_g_protocol}
\end{equation}
which is a factorized state and describes the resonator in the Fock state with $N$ photons and the QB in its ground state when the GJJ circuit is isolated, as defined in Eq.~\eqref{eq:QB_empty_full_states}. We note that, starting from $|\psi (0)\rangle $, even if all photons are absorbed simultaneously in single-photon processes, at most $N$ distinct TLSs can be excited. This choice of initial state is convenient for the numerical analysis, but more general and easier to implement initial states of the resonator, such as coherent states with average number of photons equal to $N$, have been demonstrated to lead to compatible results in conventional Dicke QBs \cite{andolina2018, andolina2019work}.
Fig.~\ref{fig:Energy_scan_g} displays the time evolution of the stored energy for various values of $\phi$, with the coupling constant fixed at $g=0.1$~\subref{fig:Energy_scan_g_0.100} and $g=0.3$~\subref{fig:Energy_scan_g_0.300}.
The stored energy is expressed in units of $N\omegar$, corresponding to the initial energy supplied by the charger.
Consequently, the ratio $E^{(g)}_{\rm B}(t)/(N\omegar)$ quantifies the fraction of energy transferred from the charger to the QB during the charging process.
Fig.~\ref{fig:Energy_scan_g_0.100} shows particular values of $\phi$ at which the charging exhibits an enhancement.
All these values of the superconducting phase difference correspond to situations where at least two pairs of units are in single-photon or two-photon resonance. However, the reverse is not always true: for instance, at $\phi = 0.46 \pi$ (pink), there are two pairs of units in a two-photon resonance condition, but no clear enhancement of the charging is observed.
This happens when the resonant units exhibit nearly perfect transmission, $\tau_n \approx 1$.
Consequently, according to Eq.~\eqref{eq:ABSs_paramg_diamg_terms}, their transverse couplings, $P_n^{x}(\phi)$ and $D_n^{x}(\phi)$, are strongly suppressed for superconducting phase differences far from $\phi = \pi$, thereby inhibiting the exchange of energy between the resonator and the QB system.
%
Moreover, the largest stored energy, $E^{(g)}_{\rm B}(t)/(N \omegar)\simeq40 \%$, is reached when there is a subset of units of the QB system whose bare energy splittings are in a two-photon resonance condition at $\phi=0.53\pi$ (orange dashed line).
Conversely, in Fig.~\ref{fig:Energy_scan_g_0.300}, the resonant processes cannot be distinguished as clearly. In this case, characterized by a larger value of the coupling, stored energies up to $E^{(g)}_{\rm B}(t)/(N \omegar)\simeq60 \%$ are generally reached at earlier times compared to the situation with $g=0.1$, and larger values are spread more uniformly across extended regions.  Nevertheless, the maximum stored energy occurs for a superconducting phase difference in the range $0.3\pi\lesssim \phi\lesssim 0.6\pi$, where two-photon resonant processes take place. In contrast to the $g=0.1$ case, the corresponding charging peaks are not well separated here.

\begin{figure*}[t]
  \centering
  \begin{subfigure}[t]{0.485\textwidth}
    \caption{}
    \begin{overpic}[width=\linewidth]{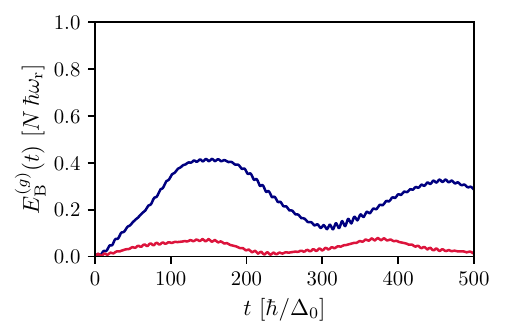}
        \put(74,56){\scriptsize
        $\begin{aligned}
            \phi/\pi&=0.53\\[-3pt]
            g&=0.1
        \end{aligned}$
        }
    \end{overpic}
    \label{fig:Energy_full_vs_no_Pz_tmax_500_phi_0.53_g_0.100}
  \end{subfigure}
  \hfill
  \begin{subfigure}[t]{0.485\textwidth}
    \caption{}
    \begin{overpic}[width=\linewidth]{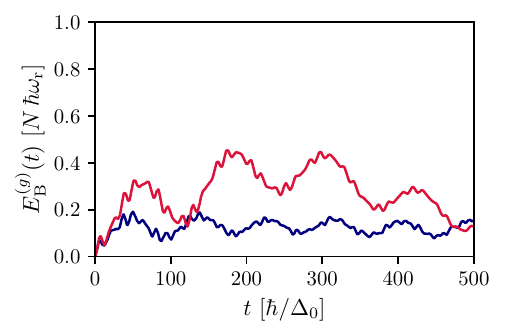}
        \put(74,56){\scriptsize
        $\begin{aligned}
            \phi/\pi&=0.76\\[-3pt]
            g&=0.1
        \end{aligned}$
        }
    \end{overpic}
    \label{fig:Energy_full_vs_no_Pz_tmax_500_phi_0.76_g_0.100}
  \end{subfigure}

  \vspace{-2.0em}

  \begin{subfigure}[t]{0.485\textwidth}
    \caption{}
    \begin{overpic}[width=\linewidth]{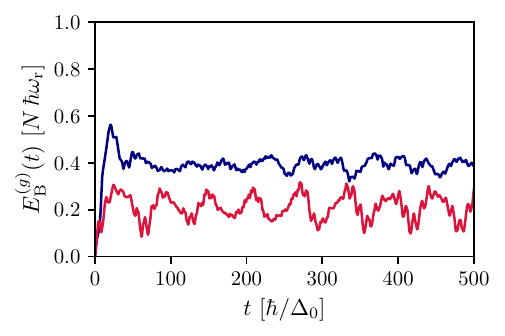}
        \put(74,56){\scriptsize
        $\begin{aligned}
            \phi/\pi&=0.53\\[-3pt]
            g&=0.3
        \end{aligned}$
        }
    \end{overpic}
    \label{fig:Energy_full_vs_no_Pz_tmax_500_phi_0.53_g_0.300}
  \end{subfigure}
  \hfill
  \begin{subfigure}[t]{0.485\textwidth}
    \caption{}
    \begin{overpic}[width=\linewidth]{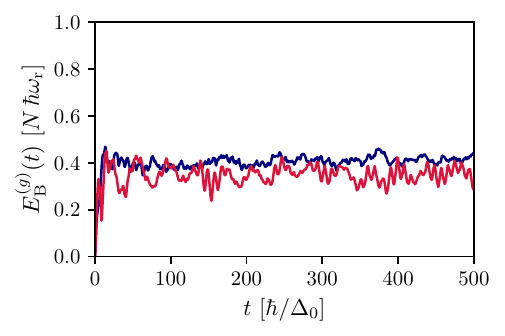}
        \put(74,56){\scriptsize
        $\begin{aligned}
            \phi/\pi&=0.76\\[-3pt]
            g&=0.3
        \end{aligned}$
        }
    \end{overpic}
    \label{fig:Energy_full_vs_no_Pz_tmax_500_phi_0.76_g_0.300}
  \end{subfigure}

  \caption{Comparison between the time evolution of the energy stored in the QB system (in units of $N\omegar$) obtained according to the complete Hamiltonian in Eq.~\eqref{eq:compact_Hamiltonian_tot} (blue solid lines) and the corresponding results computed discarding the longitudinal terms at first order in $g$ (red solid lines), i.e., the one proportional to $P_n^z(\phi)$.
  We show the results for $\phi/\pi=0.53$ \subref{fig:Energy_full_vs_no_Pz_tmax_500_phi_0.53_g_0.100}--\subref{fig:Energy_full_vs_no_Pz_tmax_500_phi_0.53_g_0.300} and for $\phi/\pi=0.76$ \subref{fig:Energy_full_vs_no_Pz_tmax_500_phi_0.76_g_0.100}--\subref{fig:Energy_full_vs_no_Pz_tmax_500_phi_0.76_g_0.300}. In panels \subref{fig:Energy_full_vs_no_Pz_tmax_500_phi_0.53_g_0.100} and \subref{fig:Energy_full_vs_no_Pz_tmax_500_phi_0.76_g_0.100} we report the case with $g=0.1$, while in the panels \subref{fig:Energy_full_vs_no_Pz_tmax_500_phi_0.53_g_0.300} and \subref{fig:Energy_full_vs_no_Pz_tmax_500_phi_0.76_g_0.300} we set $g=0.3$. In all panels, the times are reported in units of $\hbar/\deltaO$. Other parameters are $\omegar=0.75\deltaO$, $\mu_0 = 6.5\hvf/L$ and $W = 2 L$.}
  \label{fig:QB_full_vs_no Pz_tests}
\end{figure*}

\begin{figure*}[t]
  \centering
  \begin{subfigure}[t]{0.485\textwidth}
    \caption{}
        \begin{overpic}[width=\linewidth]{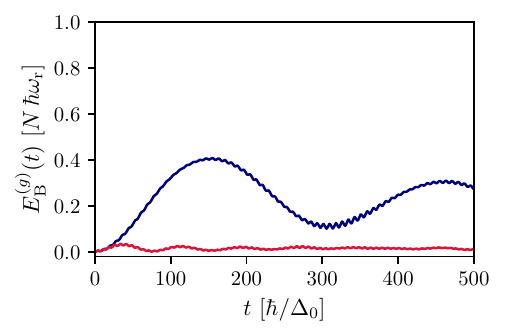}
        \put(74,56){\scriptsize
        $\begin{aligned}
            \phi/\pi&=0.53\\[-3pt]
            g&=0.1
        \end{aligned}$
        }
    \end{overpic}
    \label{fig:Energy_full_R_NR_tmax_500_g_0.100_phi_0.53}
  \end{subfigure}
  \hfill
  \begin{subfigure}[t]{0.485\textwidth}
    \caption{}
        \begin{overpic}[width=\linewidth]{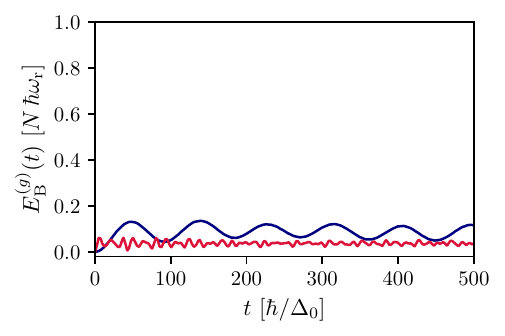}
        \put(74,56){\scriptsize
        $\begin{aligned}
            \phi/\pi&=0.76\\[-3pt]
            g&=0.1
        \end{aligned}$
        }
    \end{overpic}
    \label{fig:Energy_full_R_NR_tmax_500_g_0.100_phi_0.76}
  \end{subfigure}

  \vspace{-2.0em}

  \begin{subfigure}[t]{0.485\textwidth}
    \caption{}
        \begin{overpic}[width=\linewidth]{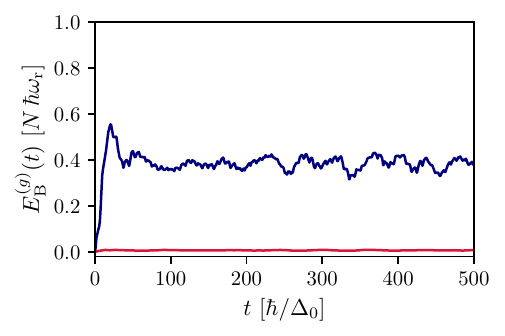}
        \put(74,56){\scriptsize
        $\begin{aligned}
            \phi/\pi&=0.53\\[-3pt]
            g&=0.3
        \end{aligned}$
        }
    \end{overpic}
    \label{fig:Energy_full_R_NR_tmax_500_g_0.300_phi_0.53}
  \end{subfigure}
  \hfill
  \begin{subfigure}[t]{0.485\textwidth}
    \caption{}
        \begin{overpic}[width=\linewidth]{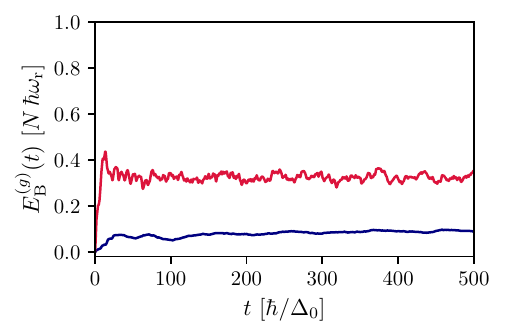}
        \put(74,56){\scriptsize
        $\begin{aligned}
            \phi/\pi&=0.76\\[-3pt]
            g&=0.3
        \end{aligned}$
        }
    \end{overpic}
    \label{fig:Energy_full_R_NR_tmax_500_g_0.300_phi_0.76}
  \end{subfigure}

  \caption{Time evolution of the energy stored in the QB system (expressed in units of $N\omegar$), separated into a near-resonant part (blue solid lines) and an off-resonant part (red solid lines), as defined in Eq.~\eqref{eq:Eg:res}. In panels \subref{fig:Energy_full_R_NR_tmax_500_g_0.100_phi_0.53} and \subref{fig:Energy_full_R_NR_tmax_500_g_0.300_phi_0.53}, we set $\phi/\pi=0.53$ and have the subset of near-resonant TLSs as $\mathcal{S} \equiv \{\nu = \pm 4\}$ for $g=0.1$ and $\mathcal{S} \equiv \{\nu =\pm 3, \pm 4\}$ for $g=0.3$. In panels \subref{fig:Energy_full_R_NR_tmax_500_g_0.100_phi_0.76} and \subref{fig:Energy_full_R_NR_tmax_500_g_0.300_phi_0.76}, we instead take $\phi/\pi=0.76$, for which the near-resonant TLSs are given by $\mathcal{S} \equiv \{\nu = \pm 1,\pm 2\}$.
  Panels \subref{fig:Energy_full_R_NR_tmax_500_g_0.100_phi_0.53} and \subref{fig:Energy_full_R_NR_tmax_500_g_0.100_phi_0.76} show the case with $g=0.1$, whereas in panels \subref{fig:Energy_full_R_NR_tmax_500_g_0.300_phi_0.53} and \subref{fig:Energy_full_R_NR_tmax_500_g_0.300_phi_0.76} we take $g=0.3$. In all panels, time is measured in units of $\hbar/\deltaO$. The remaining parameters are $\omegar=0.75\deltaO$, $\mu_0 = 6.5\hvf/L$, and $W = 2 L$.}
  \label{fig:QB_full_R_NR_componets}
\end{figure*}

To further investigate the time evolution of the stored energy, we focus on the two representative cases $\phi=0.53\pi$ (orange dashed line) and $\phi=0.76\pi$ (brown dashed line), reported in Fig.~\ref{fig:Energy_scan_g}, corresponding to a two-photon and single-photon resonance condition for some ABSs pairs,  respectively.
In the former case ($\phi=0.53\pi$), represented in Fig.~\ref{fig:Energy_static_phi_0.53}, the maximum stored energy achieved for $g=0.1$ (blue solid line) is comparable to the values observed for $g=0.3$ (green solid line). In contrast, in the second case ($\phi=0.76\pi$) shown in Fig.~\ref{fig:Energy_static_phi_0.76}, the stored energy corresponding to $g=0.1$ (blue solid line) remains smaller than the one for $g=0.3$ (green solid line) at all times.
Moreover, in both reported cases, the long-term evolution of the stored energy exhibits a suppression of the oscillations, and its average value approaches a constant energy.
To compute this average energy value, it is convenient to decompose the evolving state that describes the system in the eigenbasis $\{\ket{\mathcal{E}_j}\}$ of the total Hamiltonian $\hat{H}$, defined in Eq.~\eqref{eq:compact_Hamiltonian_tot}.

As a result, one obtains
\begin{equation}
    \begin{aligned}
    E^{(g)}_{\rm B}(t) =\sum_{j} \abs{c_j}^2\expval{\hat{H}_{\rm B}}{\mathcal{E}_j}-  {\langle \psi(0)|}{\hat{H}_{\rm B}}{|\psi(0)\rangle} + \sum_{\ell\neq j}\sum_{j} e^{-i\Omega_{\ell j}t} c^*_{\ell}c_{j} \mel{\mathcal{E}_\ell}{\hat{H}_{\rm B}}{\mathcal{E}_j}~,
    \end{aligned}
\end{equation}
where $\Omega_{\ell j}= (\mathcal{E}_j-\mathcal{E}_\ell)/\hbar$ and $c_j = \langle \mathcal{E}_j|{\psi(0)\rangle}$ denotes the $j$‑th component of the initial state in Eq.~\eqref{eq:initial_state_g_protocol}, obtained by projecting it onto the eigenbasis $\{\ket{\mathcal{E}_j}\}$, which is calculated via numerical diagonalization of $\hat{H}$.

It is straightforward to recognize that the time-independent component of $E^{(g)}_{\rm B}(t)$ is given by
\begin{equation}
\bar{E}^{(g)}_{\rm B} = \sum_{j} \abs{c_j}^2\expval{\hat{H}_{\rm B}}{\mathcal{E}_j}-  {\langle \psi(0)|}{\hat{H}_{\rm B}}{|\psi(0)\rangle}~.
    \label{eq:QB_energy_static}
\end{equation}
In the cases shown in Figs.~\ref{fig:Energy_static_phi_0.53} and~\ref{fig:Energy_static_phi_0.76}, the constant term introduced in Eq.~\eqref{eq:QB_energy_static} is represented as a dashed line, corresponding to $g=0.1$ (cyan) and $g=0.3$ (light green), respectively.

Here, for the smallest coupling constant, $g = 0.1$ (blue solid line), clear oscillations are observed, whereas for $g = 0.3$ the oscillations are less pronounced.
This aligns with the expansion of $\psiO$ in the eigenbasis $\{\ket{\mathcal{E}_j}\}$. In particular, Fig.~\ref{fig:Abs_Overlap_phi_0.53} and Fig.~\ref{fig:Abs_Overlap_phi_0.76} display the squared magnitudes of the coefficients, $|c_j|^2 = |\ip{\mathcal{E}_j}{\psi(0)}|^2$, for the chosen cases $\phi/\pi = 0.53$ and $\phi/\pi = 0.76$.

For $g=0.1$ (blue bars), the initial state can essentially be represented as a linear combination of only a small number of eigenstates of $\hat{H}$.
Accordingly, in this case, the stored energy $E^{(g)}_{\rm B}$ evolves in time as a function of a limited set of frequencies $\Omega_{\ell j}$ only, which therefore appear clearly distinguishable.
In contrast, for $g=0.3$ (green bars), the distribution of the components of the eigenstates of $\hat{H}$ is broader and poorly resolved, resulting in large frequencies $\Omega_{\ell j}$ involved in the dynamics.
As a consequence, the time evolution of $E^{(g)}_{\rm B}$ (green solid line), shown in Fig.~\ref{fig:Energy_static_phi_0.53} and Fig.~\ref{fig:Energy_static_phi_0.76}, displays a rapidly oscillating behaviour.
Furthermore, we observe that the timescale over which the energy profile $E^{(g)}_{\rm B}(t)$ settles around the stationary component $\bar{E}^{(g)}_{\rm B}(t)$ is reduced by approximately one order of magnitude compared to the $g=0.1$ case (blue solid line).

We conclude this section by discussing the role played by the longitudinal term appearing at first order in $g$ in Eq.~\eqref{eq:compact_Hamiltonian_tot}, namely the one proportional to $P_n^z(\phi)$. Microscopically, this longitudinal term accounts for the equilibrium supercurrent carried by the occupied ABS through the JJ disconnected from the LC resonator, and there is no equivalent term in the conventional Dicke Hamiltonian~\cite{dicke1954, kirton2018}. Here, we examine the numerically evaluated energy evolution for representative cases.
Fig.~\ref{fig:QB_full_vs_no Pz_tests} shows how the energy stored in the QB system evolves both in the presence (blue solid lines) and in the absence (red solid lines) of the terms $P_n^z(\phi)$.

From a comparison with the energy scheme in Fig.~\ref{fig:QB_energies}, we identify a subset of TLSs, denoted by $\mathcal{S}$, which are resonant or near-resonant with the resonator mode. More precisely, $\mathcal{S}$ contains those TLSs for which the detuning from the relevant resonance is smaller than the characteristic energy scale set by the corresponding interaction matrix element, as derived in App.~\ref{app:B}. In the parameter regime reported in Fig.~\ref{fig:Energy_full_vs_no_Pz_tmax_500_phi_0.53_g_0.100}, $\mathcal{S}$ is determined according to the two-photon resonance criterion in Eq.~\eqref{eq:app_two_photon_criterion}.

In this situation, the charging process is expected to be predominantly governed by two-photon resonance processes. From the comparison between the curves in Fig.~\ref{fig:Energy_full_vs_no_Pz_tmax_500_phi_0.53_g_0.100}, it is clear that the terms $P_n^z(\phi)$ play a supporting role in the charging process.
In Fig.~\ref{fig:Energy_full_vs_no_Pz_tmax_500_phi_0.53_g_0.300}, we keep the same phase difference but increase the coupling constant to $g=0.3$. We again find that the terms $P_n^z(\phi)$ provide a positive contribution to energy storage.

In the case shown in Fig.~\ref{fig:Energy_full_vs_no_Pz_tmax_500_phi_0.76_g_0.100}, the subset $\mathcal{S}$ is instead formed by the TLSs whose detuning from the single-photon resonance is smaller than the corresponding first-order interaction matrix element, according to Eq. \eqref{eq:app_single_photon_criterion}.

Here, the charging process is expected to be predominantly governed by single-photon resonance processes. In this regime, we observe that the terms $P_n^z(\phi)$ act as a source of disturbance in the charging dynamics: removing this contribution from the Hamiltonian improves energy storage.
Keeping the same phase difference but increasing the coupling constant to $g=0.3$, Fig.~\ref{fig:Energy_full_vs_no_Pz_tmax_500_phi_0.76_g_0.300} shows that the larger value of $g$ modifies the influence of $P_n^z(\phi)$ on energy storage, resulting in a slightly supporting contribution.

To clarify the results discussed above, in Fig.~\ref{fig:QB_full_R_NR_componets} we analyze the representative cases by separating the contribution to the stored energy into that arising from the TLSs belonging to the near-resonant subset $\mathcal{S}$, defined according to the criteria introduced above and derived in App.~\ref{app:B}, and that coming from the remaining TLSs.

Formally, we decompose the stored energy as
\begin{subequations}\label{eq:Eg:res}
\begin{align}
E^{(g)}_{\rm B}(t) &= {E^{(g){\rm \text{-}I}}_{\rm B}}(t)+{E^{(g){\rm \text{-}II}}_{\rm B}}(t)~,\\
{E^{(g){\rm \text{-}I}}_{\rm B}}(t) &= \sum_{n \in {\cal S}} \epsn\left[
\expval{\sigmaz_{n}}{\psi (t)} -  \expval{\sigmaz_{n}}{\psi(0)}\right]~,\\
{E^{(g){\rm \text{-}II}}_{\rm B}}(t) &= \sum_{n \notin {\cal S}} \epsn\left[
\expval{\sigmaz_{n}}{\psi (t)} -  \expval{\sigmaz_{n}}{\psi(0)}\right]~.
\end{align}
\end{subequations}
Here, as indicated, ${E^{(g){\rm \text{-}I}}_{\rm B}}(t)$ (blue solid lines) represents the near-resonant contribution, corresponding to the energy stored in the TLSs belonging to $\mathcal{S}$, while ${E^{(g){\rm \text{-}II}}_{\rm B}}(t)$ (red solid lines) represents the off-resonant contribution, associated with all the remaining TLSs.
In the case $\phi/\pi = 0.53$, the TLSs belonging to $\mathcal{S}$ satisfy the near two-photon resonance criterion in Eq.~\eqref {eq:app_two_photon_criterion}.

In this regime, we observe that the near-resonant contribution ${E^{(g){\rm \text{-}I}}_{\rm B}}(t)$ overwhelmingly dominates over ${E^{(g){\rm \text{-}II}}_{\rm B}}(t)$. This indicates that, in this parameter regime, the charging processes can be effectively described in terms of two-photon resonant processes over a broad range of $g$ values.

Otherwise, when the phase difference is fixed at $\phi/\pi=0.76$, the TLSs belonging to $\mathcal{S}$ satisfy the near single-photon resonance criterion in Eq. \eqref{eq:app_single_photon_criterion}.

In this regime, the near-resonant contribution ${E^{(g){\rm \text{-}I}}_{\rm B}}(t)$  is slightly larger than the remaining contribution ${E^{(g){\rm \text{-}II}}_{\rm B}}(t)$ for $g=0.1$. However, when the coupling increases to $g=0.3$, the off-resonant contribution becomes dominant. This shows that in this parameter range the charging dynamics can be accurately captured in terms of single-photon resonant processes only for sufficiently small couplings $g \ll 1$.

To better understand these representative cases, we perform a perturbative analysis. Although this method is approximate and limited in its validity, it allows us to obtain qualitative insight for the interpretation of the numerical results obtained by numerically computing the evolution governed by the full
Hamiltonian in Eq.~\eqref{eq:compact_Hamiltonian_tot}.
For convenience, we work in the interaction picture~\cite{sakurai_book}, where the total Hamiltonian can be written as
\begin{equation}
  \tilde{H}(t)=\tilde{H}_0 + \tilde{V}(t)~,
\end{equation}
with
\begin{subequations}
\begin{align}
\tilde{H}_0&=\hat{H}_{\rm r}+ \hat{H}_{\rm B}~,\\
\tilde{V}(t)&=\tilde{V}_1(t)+\tilde{V}_2(t)~,\\
\tilde{V}_1(t)&=\bar{g}(t) (\A e^{-i \omega_{\rm r} t} + \Ac e^{i \omega_{\rm r} t})\sum_{n=1}^{N}\left[P^z_{n} \hat{\sigma}^z_{n}+P^x_{n} \left(\hat{\sigma}^-_{n} e^{-i 2\varepsilon_n(\phi)t/\hbar}+\hat{\sigma}^+_{n} e^{i 2\varepsilon_n(\phi)t/\hbar}\right)\right]
~, \label{eq:V1_int_picture}\\
\tilde{V}_2(t)&=\frac{\bar{g}(t) ^2}{2} (\A e^{-i \omega_{\rm r} t} + \Ac e^{ i \omega_{\rm r} t})^2\sum_{n=1}^{N}\left[D^z_{n} \hat{\sigma}^z_{n}+D^x_{n} \left(\hat{\sigma}^-_{n} e^{-i 2\varepsilon_n(\phi)t/\hbar}+\hat{\sigma}^+_{n} e^{i 2\varepsilon_n(\phi)t/\hbar}\right)\right]\label{eq:V2_int_picture}~.
\end{align}
\label{eq:Hamiltonian_int_picture}
\end{subequations}
It is worth noting that Eqs.~\eqref{eq:V1_int_picture}--\eqref{eq:V2_int_picture} show that the dimensionless coupling constant $g$ alone is not the appropriate expansion parameter of the perturbative analysis. Indeed, the actual relevant small parameters are $g P^x_n(\phi)$, $g P^z_n(\phi)$ for single-photon processes, and
$g^2 D^x_n(\phi)/2$, $g^2D^z_n(\phi)/2$ for two-photon processes, which depend on $g$, on the superconducting phase
difference $\phi$ and on the distribution of the transmission probabilities $\{\tau_n\}$. In the cases considered in this study, typical values are $\abs{P^z_n(\phi)/\Delta_0}\lesssim 0.5$, $\abs{P^x_n(\phi)/\Delta_0}\lesssim 0.5$, $\abs{D^z_n(\phi)/\Delta_0}\lesssim 0.25$ and $\abs{D^x_n(\phi)/\Delta_0}\lesssim 0.15$. Therefore, the effective couplings are reduced with respect to the bare values of $g$.
In what follows, for the sake of simplicity, we continue to refer to the contributions in Eq.~\eqref{eq:V1_int_picture} as first order in $g$, and to those in Eq.~\eqref{eq:V2_int_picture} as second order in $g$.

Using the time-dependent perturbation theory up to second order in $g$, the state of the system can be written as
\begin{equation}
\ket{\tilde{\psi}(t)}\approx \ket*{\tilde{\psi}^{(0)}}+\ket*{\tilde{\psi}^{(1)}(t)}+\ket*{\tilde{\psi}^{(2)}(t)}~,
\end{equation}
\label{eq:State_pert_approx}
where
\begin{subequations}
\begin{align}
\ket*{\tilde{\psi}^{(0)}}&=\ket{\psi(0)}~,\\
\ket*{\tilde{\psi}^{(1)}(t)}&=\int_0^t \frac{d t_1}{i \hbar} \tilde{V}_1(t_1) \ket{\psi(0)}~,\\
\ket*{\tilde{\psi}^{(2)}(t)}&=\int_0^t \frac{d t_1}{i \hbar} \left[\tilde{V}_2(t_1) + \tilde{V}_1(t_1)\int_0^{t_1} \frac{d t_2}{i \hbar} \tilde{V}_1(t_2)\right]
\ket{\psi(0)}~.
\end{align}
\label{eq:State_pert_evolution}
\end{subequations}

We now identify the perturbative contributions that are most relevant for the charging dynamics. The first-order correction contains terms of the form $\A \sigma_n^+$ and $\Ac \sigma_n^-$, each weighted by a factor $P_n^x(\phi)$. These correspond to single-photon resonant processes, which in the rotating frame evolve with the detuning $\omega_{\rm r}-2\varepsilon_n(\phi)/\hbar$.

Exactly at the single-photon resonance, these terms become time-independent. More generally, they play a key role in the charging dynamics for the subset $\mathcal{S}$ of TLSs satisfying the near-resonance criterion in Eq. \eqref{eq:app_single_photon_criterion}, where the detuning is smaller than the corresponding first-order interaction matrix element.

These terms are also present in the conventional Dicke model~\cite{ferraro2018high}, and the presence of the terms $P_n^z(\phi)$ does not introduce an additional single-photon resonant mechanism.
The second-order correction contains terms $\A{}^2\sigma_n^{+}$ and $\Ac{}^2\sigma_n^{-}$, weighted by $D_n^x$, corresponding to two-photon resonant processes with detuning $2[\omega_{\rm r}-\varepsilon_n(\phi)/\hbar]$. Exactly at the two-photon resonance condition, these contributions become time-independent. The presence of terms $P_n^z(\phi)$ together with $P_m^x(\phi)$ generates additional terms of the form $\A{}^2\sigma_m^{+}\sigma_n^{z}$ and $\Ac{}^2\sigma_m^{-}\sigma_n^{z}$, each weighted by $P_m^x(\phi)P_n^z(\phi)$, representing further resonant two-photon mechanisms.

Moreover, the terms $P_n^z(\phi)$ generate time-independent contributions of order $g^2$. In particular, terms of the form $(2\A^\dagger\A+1)\hat{\sigma}_n^z\hat{\sigma}_m^z$, weighted by $P_n^z(\phi)P_m^z(\phi)$ with $n,m=1,\dots,N$, appear and correspond to $N^2$ photon-mediated longitudinal couplings between ABSs. In addition, $\tilde V_2$ produces  $N$ terms of the type $(2\A^\dagger\A+1)\hat{\sigma}_n^z$, weighted by $D_n^z(\phi)$.
During the charging process, these contributions of order $g^2$  lead to effective state-dependent energies $\varepsilon_n(\phi)+g^2\Xi_n(t)$, where $\Xi_n(t)$ depends on quantities such as $\langle\A^\dagger\A\rangle$ and $\langle\hat{\sigma}_m^z\rangle$. As a consequence, the detuning becomes dynamically renormalized during the charging process. An ensemble of TLSs that is initially tuned to the single-photon or two-photon resonance condition can therefore be shifted out of resonance as the photon population and ABS occupations evolve, thus hampering the charging dynamics.

In summary, when a subset of TLSs is tuned to a two-photon resonance, the terms $P_n^z(\phi)$ have a beneficial effect because it generates additional resonant processes. By contrast, when the system is tuned to a single-photon resonance, the terms $P_n^z(\phi)$ do not produce additional resonant contributions. Its dominant effect is, instead, to induce a time-dependent detuning, which tends to hinder the charging dynamics.
We note that the time-dependent detuning induced by the terms $P_n^z(\phi)$ is also present in the two-photon resonance regime. In that setting, however, its impact is relatively weaker, because the additional resonant processes it generates provide a dominant contribution that compensates for the detrimental effect of the time-dependent detuning.
This interpretation explains the results obtained when the phase difference is set at $\phi=0.53\pi$ (two-photon resonance) over a wide range of values of $g$, and when $\phi=0.76\pi$ (single-photon resonance) for small $g$, such as $g=0.1$, where the charging process is mainly governed by the resonant TLSs. Finally, in the case shown in Fig.~\ref{fig:Energy_full_vs_no_Pz_tmax_500_phi_0.76_g_0.300} and Fig.~\ref{fig:Energy_full_R_NR_tmax_500_g_0.300_phi_0.76}, where the phase difference is fixed at $\phi=0.76\pi$ (single-photon resonance) and the coupling is relatively large, $g=0.3$, off-resonant contributions become important. In particular, even though the system is nominally tuned for single-photon resonance, the relatively large value of $g$ also activates two-photon processes. In this regime, the presence of the terms $P_n^z(\phi)$ improves the efficiency of these processes, so its overall effect becomes beneficial, in contrast to the small $g$ regime where two-photon processes remain marginal.
Additional details on the role of the terms $P_n^{z}(\phi)$ are provided in App.~\ref{app:A}.
\subsection{Single transmission probability}
\label{subsec:Single transmission probability}

\begin{figure*}[t]
  \centering
  \begin{subfigure}[t]{0.485\textwidth}
     \caption{}
     \includegraphics[width=\textwidth]{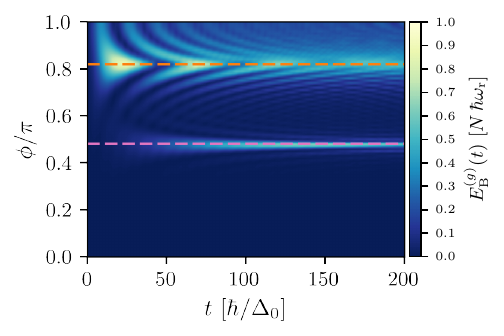}
     \label{fig:Equal_energy_scan_g_0.100}
  \end{subfigure}
  \hfill
    \begin{subfigure}[t]{0.485\textwidth}
      \caption{}
     \includegraphics[width=\textwidth]{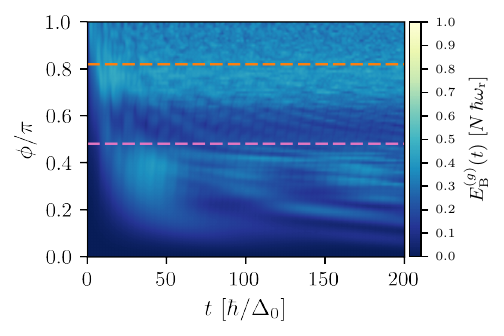}
     \label{fig:Equal_energy_scan_g_0.300}
  \end{subfigure}

  \caption{Time evolution of the energy stored in the QB system (in units of $N\omegar$), evaluated for $g=0.1$ \subref{fig:Equal_energy_scan_g_0.100} and $g=0.3$ \subref{fig:Equal_energy_scan_g_0.300}
  as a function of the superconducting phase difference $\phi$. The two values of the superconducting phase difference $\phi/\pi=0.82$ (orange dashed line) and $\phi/\pi=0.48$ (pink dashed line) correspond to single-photon and double-photon resonance conditions, respectively.
  In all panels, the times are reported in units of $\hbar/\deltaO$.
  Other parameters are $\tau_{\rm eff} = 0.93$, $N = 8$, $\hbar\omega_{\rm r} = 0.75 \deltaO$. }
  \label{fig:Equal_energy_scan_g_small}
\end{figure*}

In this section,  to explicitly assess the sensitivity of our results to the specific distribution of $\{\tau_n\}$ given by Eq.~\eqref{eq:graphene_normal_phase_transmission}, we compare the analysis discussed above with a simplified effective model in which all $N$ conduction channels of the short hybrid JJ are assumed to share the same transport characteristics. In other words, each channel is characterized by the same effective transmission probability, $\tau_{\rm eff}$, resulting in a uniform distribution of the ABSs splitting energies.
In particular, under this simplified description, we examine the charging protocol discussed earlier and, by comparing with the results of the previous section, we highlight the impact of a non‑flat distribution of ABSs splitting energies.

The reason behind this approach lies in the fact that, frequently, in experimental studies of short hybrid JJ devices, ABSs are described by introducing a single effective transmission probability, $\tau_{\rm eff}$, which is extracted by fitting the CPR measured across the junction\cite{iorio2023}.

Similarly, in this study, to determine this single effective transmission probability we compute the CPR of the junction, $I(\phi)$, via Eq.~\eqref{eq:CPR_hybrid_JJ}, which sums the contributions of all $N$ conduction channels each characterized by its transmission probability $\tau_n$ evaluated by keeping the same Fermi level $\mu_0=6.5 \hbar v_{\rm F}/L$ and junction aspect ratio $W/L=2$, as in Fig.~\ref{fig:QB_energies}. Then, by fitting $I(\phi)$ with the expression for the CPR obtained from Eq.~\eqref{eq:CPR_hybrid_JJ} upon replacing every $\tau_n$ by one common value $\tau_{\rm
eff}$, we extract a single effective transmission probability $\tau_{\rm eff}\approx0.93$ which is compatible with existing estimations obtained for highly transparent hybrid JJs~\cite{nichele2020relating,hinderling2024}.

Within this simplified picture, the total Hamiltonian assumes the form of a generalized Dicke model~\cite{dicke1954,pellegrino_prb_2014,crescente2020ultrafast}
\begin{equation}
    \begin{aligned}
    \hat{H}_{\rm gD} &=\hat{H}_{\rm r}+ \varepsilon(\phi)\sum_{n=1}^{N}\sigmaz_{n} + g (\A + \Ac)\vb{P}\cdot\sum_{n=1}^{N} \hat{\boldsymbol{\sigma}}_{n}+\frac{g^2}{2} \left(\A +\Ac\right)^2\vb{D}\cdot \sum_{n=1}^{N}\hat{\boldsymbol{\sigma}}_{n}~,
    \end{aligned}
    \label{eq:compact_equal_Hamiltonian_tot}
\end{equation}
which differs from the bare Dicke Hamiltonian due to the inclusion of a quadratic term in the operator $\A + \Ac$ and a linear term in the same operator, $\A + \Ac$, which is weighted by $P^z(\phi)$, and commutes with $\hat{H}_{\rm B}$.
Here, the quantities $\varepsilon(\phi)$, $\vb{P}$, and $\vb{D}$ are obtained from Eq.~\eqref{eq:ABSs_energy} and Eq.~\eqref{eq:ABSs_paramg_diamg_terms} by replacing $\tau_n$ with $\tau_{\rm eff}$, namely,
\begin{subequations}
\begin{align}
\varepsilon(\phi)&=\deltaO\sqrt{1-\tau_{\rm eff}\Msinq{\phi/2}}~,\\
        \vb{P}&= \left(P^{z}(\phi),\,P^{x}(\phi)\right)^{T} =\df\varepsilon(\phi)\left(1,\,-\sqrt{1-\tau_{\rm eff}}\Mtan{\frac{\phi}{2}}\right)^{T}~,\\
        \vb{D} &= \left(D^{z}(\phi),\,D^{x}(\phi)\right)^{T}=\df\varepsilon(\phi)\left(\frac{\tau_{\rm eff} +(2-\tau_{\rm eff}) \Mcos{\phi}}{2\Msin{\phi}},\,-\sqrt{1-\tau_{\rm eff}}\right)^{T}~.
\end{align}
\label{eq:Equal_ABSs_paramg_diamg_terms}
\end{subequations}

\begin{figure*}[t]
  \centering
  \begin{subfigure}[t]{0.485\textwidth}
    \caption{}
        \begin{overpic}[width=\linewidth]{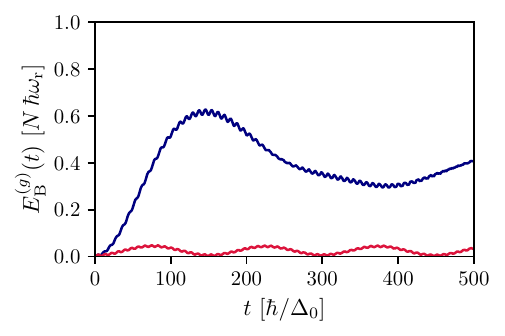}
        \put(74,56){\scriptsize
        $\begin{aligned}
            \phi/\pi&=0.48\\[-3pt]
            g&=0.1
        \end{aligned}$
        }
    \end{overpic}
    \label{fig:Energy_full_vs_no_Pz_tmax_500_phi_0.48_g_0.100}
  \end{subfigure}
  \hfill
  \begin{subfigure}[t]{0.485\textwidth}
    \caption{}
        \begin{overpic}[width=\linewidth]{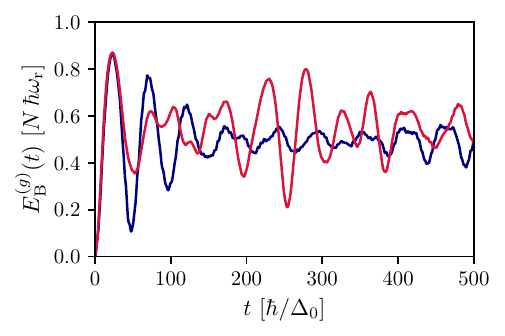}
        \put(74,56){\scriptsize
        $\begin{aligned}
            \phi/\pi&=0.82\\[-3pt]
            g&=0.1
        \end{aligned}$
        }
    \end{overpic}
    \label{fig:Energy_full_vs_no_Pz_tmax_500_phi_0.82_g_0.100}
  \end{subfigure}

  \vspace{-2.0em}

  \begin{subfigure}[t]{0.485\textwidth}
    \caption{}
        \begin{overpic}[width=\linewidth]{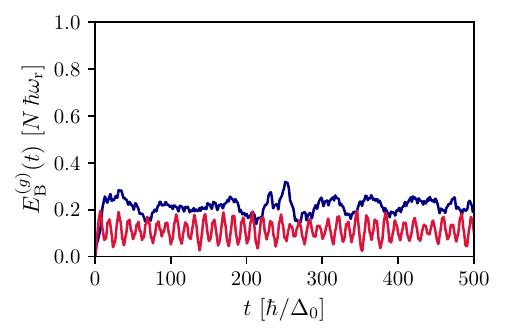}
        \put(74,56){\scriptsize
        $\begin{aligned}
            \phi/\pi&=0.48\\[-3pt]
            g&=0.3
        \end{aligned}$
        }
    \end{overpic}
    \label{fig:Energy_full_vs_no_Pz_tmax_500_phi_0.48_g_0.300}
  \end{subfigure}
  \hfill
  \begin{subfigure}[t]{0.485\textwidth}
    \caption{}
        \begin{overpic}[width=\linewidth]{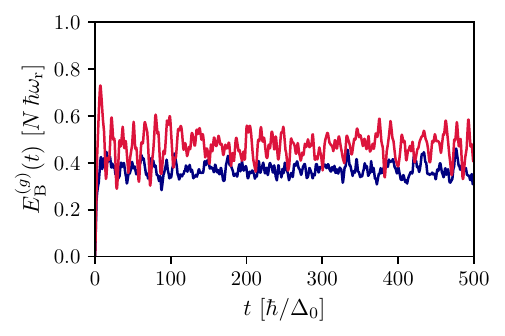}
        \put(74,56){\scriptsize
        $\begin{aligned}
            \phi/\pi&=0.82\\[-3pt]
            g&=0.3
        \end{aligned}$
        }
    \end{overpic}
    \label{fig:Energy_full_vs_no_Pz_tmax_500_phi_0.82_g_0.300}
  \end{subfigure}

  \caption{Comparison of time evolution of the energy stored in the QB system (in units of $N\omegar$) obtained according to the full Hamiltonian in Eq.~\eqref{eq:compact_Hamiltonian_tot} (blue solid lines) with the corresponding results computed by neglecting the term $P^{z}(\phi)$ (red solid lines). We show the results for $\phi/\pi=0.48$ \subref{fig:Energy_full_vs_no_Pz_tmax_500_phi_0.48_g_0.100}--\subref{fig:Energy_full_vs_no_Pz_tmax_500_phi_0.48_g_0.300} and for $\phi/\pi=0.82$ \subref{fig:Energy_full_vs_no_Pz_tmax_500_phi_0.82_g_0.100}--\subref{fig:Energy_full_vs_no_Pz_tmax_500_phi_0.82_g_0.300}. In panels \subref{fig:Energy_full_vs_no_Pz_tmax_500_phi_0.48_g_0.100} and \subref{fig:Energy_full_vs_no_Pz_tmax_500_phi_0.82_g_0.100} one has $g=0.1$, while in panels \subref{fig:Energy_full_vs_no_Pz_tmax_500_phi_0.48_g_0.300} and \subref{fig:Energy_full_vs_no_Pz_tmax_500_phi_0.82_g_0.300} one has $g=0.3$. In all panels, the times are reported in units of $\hbar/\deltaO$. Other parameters are $\tau_{\rm eff} = 0.93$, $N = 8$, $\hbar\omega_{\rm r} = 0.75 \deltaO$.}
  \label{fig:Equal_QB_full_vs_no Pz_tests}
\end{figure*}

Analogously to what has been done before, we analyze the time evolution of the energy stored in the QB, measured in units of $N \omegar$, under the charging protocol specified in Eq.~\eqref{eq:g_protocol}.
Here, exploiting the fact that now $\hat{S}^2$ is a conserved quantity, namely $\comm*{\hat{S}^2}{\hat{H}_{\rm gD}(t)}= 0$, we can restrict the computation to a smaller invariant $(N+1) \times (N+1)$ subspace for the ABSs sector, thereby reducing the computational cost required to study the time evolution of the QB system.
Moreover, we use as the initial state $|\psi(0)\rangle$, previously defined in Eq.~\eqref{eq:initial_state_g_protocol}, where the matter part  $\ket{\psidw}$ belongs to the Hilbert subspace associated with the maximal eigenvalue of the operator $\hat{S}^2$.

Fig.~\ref{fig:Equal_energy_scan_g_small} shows the time evolution of the stored energy in the QB system (in units of $N\omegar$) for the coupling strengths $g=0.1$ \subref{fig:Equal_energy_scan_g_0.100} and $g=0.3$ \subref{fig:Equal_energy_scan_g_0.300}, as a function of the superconducting phase difference $\phi$.
The specific values $\phi/\pi=0.82$ (orange dashed line) and $\phi/\pi=0.48$ (pink dashed line) correspond to the resonance conditions for the single-photon ($2\varepsilon(\phi)=\hbar \omega_{\rm r}$) and two-photon ($\varepsilon(\phi)=\hbar \omega_{\rm r}$) processes within the effective model considered, respectively. For $g=0.1$, the charging efficiency is markedly enhanced at the values of $\phi$ that correspond to configurations in which all TLSs are tuned in resonance with single- or two-photon processes. Under single-photon resonance, this leads to a maximum stored energy of $E^{(g)}_{\rm B}(t)/(N\hbar \omega_{\rm r})\simeq86 \%$, while under the two-photon resonance, the maximum stored energy reaches $E^{(g)}_{\rm B}(t)/(N\hbar \omega_{\rm r})\simeq60 \%$.

Similarly to the case of a non-flat distribution of energy splittings, for a coupling constant $g=0.3$, stored energies up to $E^{(g)}_{\rm B}(t)/(N\hbar \omega_{\rm r})\simeq 47\%$ are generally reached at earlier times than for $g=0.1$, and the larger values are distributed more evenly over broader regions of the density plot. However, the maximum stored energy is obtained when the superconducting phase difference is within the interval $0.7\pi\lesssim \phi\lesssim 0.9\pi$.

Therefore, these results indicate that within this simplified model, for both values of the coupling constant $g$ the charging process is enhanced in those energy configurations where the ABS splitting energy $2 \varepsilon(\phi)$ is confined to an energy window centered around $\hbar\omega_{\rm r}$ (single-photon resonance).

As in the case of a non-flat distribution of TLSs splitting energies, we analyze the impact of the term $P^z(\phi)$.
Figs.~\ref{fig:Energy_full_vs_no_Pz_tmax_500_phi_0.48_g_0.100}--\subref{fig:Energy_full_vs_no_Pz_tmax_500_phi_0.48_g_0.300} show the evolution of the stored energy for a fixed value $\phi/\pi=0.48$, which corresponds to tuning the TLSs to the two-photon resonance, considering the cases $g=0.1$ and $g=0.3$, respectively.
For both values of $g$, we observe that the term $P^z(\phi)$ plays a beneficial role in the charging processes, in agreement with the earlier discussion, where we noted that, together with the term $P^x(\phi)$, it introduces additional mechanisms for the creation/disruption of two-photon processes that de-excite/excite a TLS.
Figs.~\ref{fig:Energy_full_vs_no_Pz_tmax_500_phi_0.82_g_0.100}--\subref{fig:Energy_full_vs_no_Pz_tmax_500_phi_0.82_g_0.300} show the evolution of the stored energy for a fixed value $\phi/\pi=0.82$, which corresponds to tuning the TLSs to the single-photon resonance, for the two coupling strengths $g=0.1$ and $g=0.3$, respectively. In contrast to the case of a non-flat distribution of TLS splitting energies, we find that, for both values of $g$, the contribution $P^z(\phi)$ hinders the charging processes. In this case, all TLSs are tuned to satisfy the single-photon resonance condition, and the resulting behavior aligns fully with what is shown in Figs. \ref{fig:Energy_full_R_NR_tmax_500_g_0.100_phi_0.76}--\subref{fig:Energy_full_R_NR_tmax_500_g_0.300_phi_0.76}. There, for a non-flat distribution of TLS splitting energies, the blue solid lines select the charging dynamics corresponding to the subgroup of TLSs that are tuned to the single-photon condition.
A comparison between the exact model, in which the ABS energy spectrum and the effective light–matter couplings reflect the non-uniform distribution of transmission probabilities, and the simplified model with a flat distribution reveals that our main results are insensitive to this approximation. In both descriptions, the enhancement of energy storage due to two-photon processes and the role of the longitudinal coupling in the charging dynamics remain qualitatively unchanged. This indicates that the observed effects do not depend on the specific form of the distribution $\{\tau_n\}$, but rather on the light–matter coupling mechanism.

\subsection{Alternative charging protocol: Superconducting phase difference time modulation}
\begin{figure*}[t]
  \centering
  \begin{subfigure}[t]{0.485\textwidth}
    \caption{}
    \includegraphics[width=1.\textwidth]{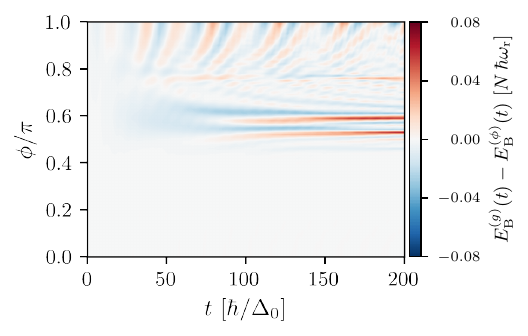}
    \label{fig:Energy_scan_comparison_g_0.100}
  \end{subfigure}
  \hfill
  \begin{subfigure}[t]{0.485\textwidth}
    \caption{}
    \includegraphics[width=1.\textwidth]{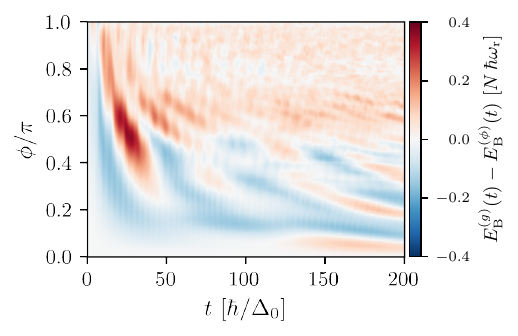}
    \label{fig:Energy_scan_comparison_g_0.300}
  \end{subfigure}

  \vspace{-2em}

  \begin{subfigure}[t]{0.485\textwidth}
    \caption{}
        \begin{overpic}[width=\linewidth]{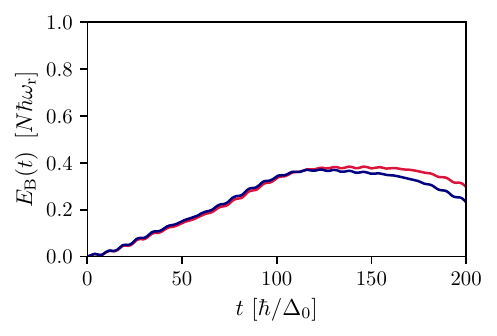}
        \put(74,57){\scriptsize
        $\begin{aligned}
            \phi/\pi&=0.53\\[-3pt]
            g&=0.1
        \end{aligned}$
        }
    \end{overpic}
    \label{fig:Energy_vs_t_protocol_comparison_phi_0.530_g_0.100}
  \end{subfigure}
  \hfill
  \begin{subfigure}[t]{0.485\textwidth}
    \caption{}
        \begin{overpic}[width=\linewidth]{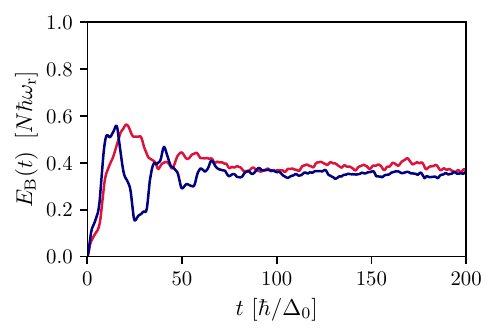}
        \put(74,57){\scriptsize
        $\begin{aligned}
            \phi/\pi&=0.53\\[-3pt]
            g&=0.3
        \end{aligned}$
        }
    \end{overpic}
    \label{fig:Energy_vs_t_protocol_comparison_phi_0.530_g_0.300}
  \end{subfigure}
  \caption{ \subref{fig:Energy_scan_comparison_g_0.100}--\subref{fig:Energy_scan_comparison_g_0.300} Time evolution of the difference between the energies stored in the QB system (in units of $N\omegar$) by employing the charging protocols in Eq.~\eqref{eq:g_protocol}, $E^{(g)}_{\rm B}$, and Eq.~\eqref{eq:phi_protocol}, $E^{(\phi)}_{\rm B}$, respectively.
  \subref{fig:Energy_vs_t_protocol_comparison_phi_0.530_g_0.100}--\subref{fig:Energy_vs_t_protocol_comparison_phi_0.530_g_0.300}
  Comparison of the time evolution of the energy stored in the QB system (in units of $N\omegar$), evaluated for $\phi/\pi = 0.53$ (two-photon resonance condition), employing the charging protocols based on coupling strength (red solid line) and based on superconducting phase difference (blue solid line), respectively.
  \subref{fig:Energy_scan_comparison_g_0.100}--\subref{fig:Energy_vs_t_protocol_comparison_phi_0.530_g_0.100} illustrate the case $g=0.1$, while \subref{fig:Energy_scan_comparison_g_0.300}--\subref{fig:Energy_vs_t_protocol_comparison_phi_0.530_g_0.300} the case $g=0.3$.
  In all panels, the times are reported in units of $\hbar/\deltaO$. Other parameters are $\omegar=0.75\deltaO$, $\mu_0 = 6.5\hvf/L$ and $W = 2 L$.}
  \label{fig:Energy_scan_comparison}
\end{figure*}

Although the protocol in Eq.~\eqref{eq:g_protocol} has been adopted as a reasonable choice in several theoretical studies~\cite{ferraro2018high, andolina2019work,razzoli2025cyclic}, its actual implementation in scalable architectures remains challenging~\cite{quach2022micro,elghaayda2025sqb}. As an alternative, we propose a second charging protocol that exploits the dependence of the ABS spectrum on the superconducting phase difference.
Here, we assume a sudden quench of the superconductive phase difference across the JJ.
This can be obtained by controlling the external magnetic fluxes in cQED setups~\cite{bryon2023}, or by a pulsed voltage drop described as $V(t)=[\hbar \phi/(2e)][\delta(t) - \delta(t-t_{\rm c})]$~\cite{GarciaRipoll_2022}.
Therefore, in the following charging protocol, we implement in Eq.~\eqref{eq:compact_Hamiltonian_tot} the substitution
\begin{equation}
    \phi \to \bar{\phi}(t) = \phi\,\Theta(t)\Theta(t_{\rm c}-t)~.
    \label{eq:phi_protocol}
\end{equation}
Here, the applied modulation enables an effective coupling (or decoupling) between the QB and the charger.
Unlike approaches based on modulating coupling constants~\cite{sete2021,campbell2023,heunisch2023}, the present scheme does not require quantum couplers to actively control interactions between circuit components~\cite{haller2022,monroe2024}. This should reduce hardware complexity and may open a promising new direction for the design and investigation of solid-state QBs.

In this alternative charging protocol scenario, before the charging process starts, one has $\phi=0$ and the total system is described by the initial Hamiltonian
\begin{equation}
    \begin{aligned}
    \hat{H}_0 &=\hbar\omega_r\left(\Ac\A +\frac{1}{2}\right)+ \Delta_0 \sum_{n}\,\sigmaz_{n}+\frac{g^2}{2} \left(\A + \Ac\right)^2\left(-\frac{\deltaO}{4}\sum_{n}\tau_n\,\sigmaz_{n} \right)~.
    \end{aligned}
    \label{eq:initial_Hamiltonian_phi}
\end{equation}
We recall that for the specific case of GJJ, one replaces the label $n$ with the composite index $\nu\equiv(\zeta,m)$. Because the operators $\sigmaz_{n}$ are conserved quantities, i.e., $\comm*{\hat{H}_0}{\sigmaz_{n}}=0$, the Hamiltonian in Eq.~\eqref{eq:initial_Hamiltonian_phi} takes a block-diagonal form when expressed in the common eigenbasis of the set $\{\sigmaz_{n}\}$.
If we consider as the initial state a generic factorized eigenstate of all $\sigmaz_{n}$, the time evolution remains confined within the corresponding invariant subspace. Therefore, each operator can be replaced by its eigenvalue $s^z_n = \pm1$, such that Eq.~\eqref{eq:initial_Hamiltonian_phi} reduces to an effective quadratic bosonic Hamiltonian parametrized by the initial configuration $\{s^z_n\}$, which reads as follows
\begin{equation}
    \begin{aligned}
    \hat{H}_0 =  \hbar\omega_r\left(\Ac\A +\frac{1}{2}\right)+\varepsilon_0 + g_{\rm D}\left(\A + \Ac\right)^2~,
    \end{aligned}
    \label{eq:initial_Hamiltonian_phi_spin_set}
\end{equation}
where $\varepsilon_0 = \deltaO\sum_{n}s^{z}_{n}$ is the total initial energy of the QB and $g_{\rm D} = - (g^2\deltaO/8)\sum_{n}\tau_n s^{z}_{n}$.
This Hamiltonian can be brought into diagonal form via a Bogoliubov transformation, resulting in
\begin{equation}
    \begin{aligned}
        \hat{H}_{0} &= \hbar\lambda\omega_r\left(\Bc\B +\frac{1}{2}\right)+ \varepsilon_0~,\\
    \end{aligned}
    \label{eq:initial_diagonal_Hamiltonian_phi_protocol}
\end{equation}
where $\B  = \Mcosh{x}\A + \Msinh{x}\Ac$ is a bosonic annihilation operator that acts on a new vacuum state as $\B\ket{0}_{\rm b} = 0$, while $x = \ln\sqrt{\lambda}$ with $\lambda = \left[1 + 4g_{\rm D}/\omegar\right]^{1/2}$.
In complete analogy to Eq.~\eqref{eq:initial_state_g_protocol}, here we consider the following initial state for the charging
\begin{equation}
    \ket{\chi(0)} = \ket{N}_{\rm b}\otimes\ket{\psidw}~,
    \label{eq:initial_state_phi_protocol}
\end{equation}
where $\ket{N}_{\rm b}=(1/\sqrt{N!}) \hat{b}^{\dagger}{}^N\ket{0}_{\rm b} $ is a Fock state that contains $N$ excitations for the new bosonic mode, while the state $\ket{\psidw}$ corresponds to the initial configuration $\{s^z_n = -1\}$.
By expressing the total Hamiltonian that characterizes the charging process in terms of the new creation (annihilation) operators $\Bc$ ($\B$), we obtain
\begin{equation}
    \begin{aligned}
    \hat{H} =  \hbar\lambda\omega_r\left(\Bc\B +\frac{1}{2}\right) + \sum_{n=1}^{N}\epsn\,\sigmaz_{n} + g_{\lambda}(\B + \Bc)\sum_{n=1}^{N}\vb{P}_{n}\cdot \hat{\boldsymbol{\sigma}}_{n}+\frac{ g_{\lambda}^2}{2} \left(\B +\Bc\right)^2\sum_{n=1}^{N}\left(\vb{D}_{n}\cdot \hat{\boldsymbol{\sigma}}_{n} + D_{0,n}^{z}\right)~,
    \end{aligned}
    \label{eq:compact_Hamiltonian_phi_protocol}
\end{equation}
where $g_\lambda = g/\sqrt{\lambda}$ is the renormalized coupling constant and $D_{0,n}^{z} = D_{n}^{z}(\phi=0)$, according to Eq.~\eqref{eq:ABSs_paramg_diamg_terms}.

Figs.~\ref{fig:Energy_scan_comparison_g_0.100}-\subref{fig:Energy_scan_comparison_g_0.300} show how the quantity $E^{(g)}_B - E^{(\phi)}_B$, i.e.,
the difference between the energy stored in the QB system using the
coupling-strength protocol and that obtained with the
superconducting-phase-difference protocol, evolves as a function of
$\phi$, for two coupling strengths, $g=0.1$ and $g=0.3$, respectively.

For $g=0.1$, the difference between the two protocols
typically remains well below $0.1\,N\hbar\omega_{\rm r}$ over most of the range of $\phi$, as shown in  Fig.~\ref{fig:Energy_scan_comparison_g_0.100}. The largest discrepancies are found in the vicinity of the two-photon
resonance conditions identified in Fig.~\ref{fig:Energy_scan_g_0.100} (orange and red dashed
lines), where the charging dynamics is dominated by two-photon resonance process. Despite this, by focusing on  the time evolution
of the stored energy at $\phi/\pi=0.53$, which represents the two-photon resonance condition discussed in Sec.~\ref{sec:results}, Fig.~\ref{fig:Energy_vs_t_protocol_comparison_phi_0.530_g_0.100} shows that the
coupling-strength protocol $E^{(g)}_B$ (red solid line) and the
phase-difference protocol $E^{(\phi)}_B$ (blue solid line) reach comparable maximum stored energies, while their difference only becomes appreciable at longer times.

Conversely, for $g=0.3$, the discrepancy between the two
protocols becomes more pronounced, reaching up to $0.4\,N\hbar\omega_{\rm
r}$ in the region around the two-photon resonance conditions, as shown in Fig.~\ref{fig:Energy_scan_comparison_g_0.300}. Furthermore, at
$\phi/\pi=0.53$, Fig.~\ref{fig:Energy_vs_t_protocol_comparison_phi_0.530_g_0.300} shows that the time evolution of the stored
energy for the coupling-strength protocol (red solid line) and the
phase-difference protocol (blue solid line) are distinguishable already at early times. Nevertheless, also in this coupling scenario, the maximum stored energy reached by the two protocols remains comparable. Notably, the phase-difference protocol exhibits faster oscillations of the stored energy during the initial stage of the dynamics, and it is precisely this fact that accounts for the sizable discrepancy between the two protocols reported in  Fig.~\ref{fig:Energy_scan_comparison_g_0.300}. Therefore, for both values of $g$, the two protocols present a comparable performance in terms of the maximum energy storage they can achieve, but not in terms of the overall time development.

We note that the sudden variation of the corresponding control parameter requires an external energy input in both protocols. In either case, the Hamiltonian term activated by the quench does not commute with the corresponding pre-quench Hamiltonian~\cite{Campaioli24}.

Given that both protocols are characterized by a comparable stored energy, particularly in the small coupling regime, and allow for an efficient exploitation of two-photon processes compared with single-photon ones, the selection between them can primarily rely on considerations dictated by practical implementation and experimental constraints.

\section{Conclusions}
In this work, we investigated the possible implementation of a Dicke-like QB based on a superconducting circuit architecture. The setup features an LC resonator that is inductively coupled to a superconducting loop containing a two-dimensional superconductor–semiconductor JJ. Focusing on the case of a finite-width GJJ in the short-junction limit, we described the ABSs within the link, each associated with a transport channel, as an ensemble of TLSs coupled with the photons into the resonator.

First, we examined the charging performance of this device using a standard protocol that is based on the energy transfer from the photons to the TLSs via a sudden change of the inductive coupling strength. In the weak-coupling regime, our findings indicate that when the system is tuned in such a way that a number of TLSs satisfy a two-photon resonance condition, the charging process is improved with respect to the cases where the single-photon resonance is achieved.

This behavior is due to the presence of a longitudinal linear coupling term in the interaction Hamiltonian, which originates microscopically from the coupling between the LC flux and the supercurrent passing through the junction. Our analysis showed that this term, which is absent in the conventional Dicke model, can generate additional two-photon resonant processes that assist the charging dynamics. These conclusions were further corroborated by comparison with an effective model in which all the TLSs composing the QB have the same energy splittings.

Finally, we introduced an alternative charging protocol based on the sensitivity of the ABSs spectrum with respect to the superconducting phase difference. Notably, when this phase difference is abruptly changed through a sudden quench, the resulting energy storage behavior is comparable to that obtained with the coupling-based protocol, highlighting how the flexibility of this platform can adapt to possible experimental limitations.
In both protocols, the Hamiltonian term activated by the sudden quench does not commute with the pre-quench Hamiltonian, and an external energy input is generally required.

Our theoretical analysis shows that the proposed device can serve as an efficient and versatile implementation of a QB in solid-state superconducting platforms. This could provide a firm theoretical basis for the future implementation of solid-state QB based quantum computer architectures such as the one proposed in Ref.~\cite{Kurman25}, which appear very promising in improving the energy efficiency of quantum computing and in mitigating the wiring problem, one of the greatest limitations in the scaling of quantum computers.

\begin{appendices}
	\appendix
	\numberwithin{equation}{section}
	\renewcommand\thefigure{\thesection.\arabic{figure}}
	\setcounter{figure}{0}

\section{Effect of longitudinal terms, linear in the coupling constant, on the charging process}\label{app:A}

\begin{figure*}[t]
  \centering
  \begin{subfigure}[t]{0.48\textwidth}
    \caption{}
        \begin{overpic}[width=\linewidth]{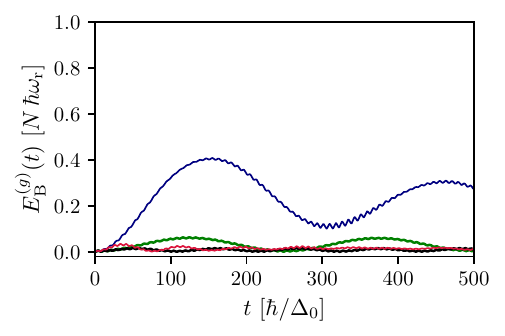}
        \put(74,56){\scriptsize
        $\begin{aligned}
            \phi/\pi&=0.53\\[-3pt]
            g&=0.1
        \end{aligned}$
        }
    \end{overpic}
    \label{fig:Energy_full_R_NR_vs_no_Pz_tmax_500g_0.100_phi_0.53}
  \end{subfigure}
  \hfill
  \begin{subfigure}[t]{0.48\textwidth}
    \caption{}
        \begin{overpic}[width=\linewidth]{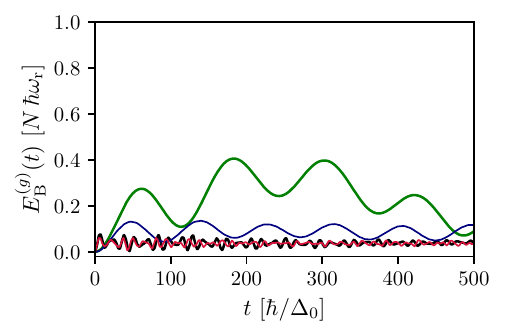}
        \put(74,56){\scriptsize
        $\begin{aligned}
            \phi/\pi&=0.76\\[-3pt]
            g&=0.1
        \end{aligned}$
        }
    \end{overpic}
    \label{fig:Energy_full_R_NR_vs_no_Pz_tmax_500g_0.100_phi_0.76}
  \end{subfigure}

  \vspace{-2em}

  \begin{subfigure}[t]{0.48\textwidth}
    \caption{}
        \begin{overpic}[width=\linewidth]{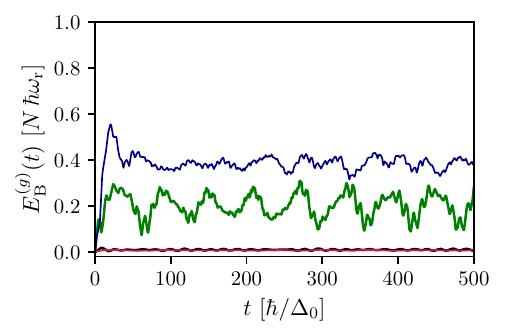}
        \put(74,56){\scriptsize
        $\begin{aligned}
            \phi/\pi&=0.53\\[-3pt]
            g&=0.3
        \end{aligned}$
        }
    \end{overpic}
    \label{fig:Energy_full_R_NR_vs_no_Pz_tmax_500g_0.300_phi_0.53}
  \end{subfigure}
  \hfill
  \begin{subfigure}[t]{0.48\textwidth}
    \caption{}
        \begin{overpic}[width=\linewidth]{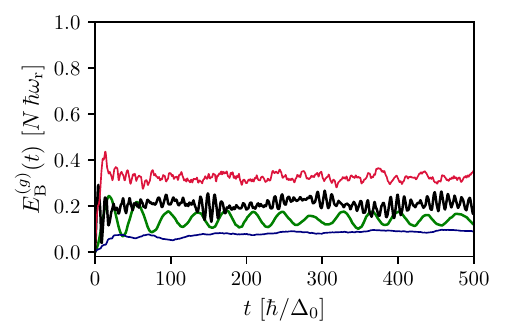}
        \put(74,56){\scriptsize
        $\begin{aligned}
            \phi/\pi&=0.76\\[-3pt]
            g&=0.3
        \end{aligned}$
        }
    \end{overpic}
    \label{fig:Energy_full_R_NR_vs_no_Pz_tmax_500g_0.300_phi_0.76}
  \end{subfigure}

  \caption{Time evolution of the stored energy in the QB system (in units of $N\omegar$), computed by neglecting the terms $P_n^{z}(\phi)$ in Eq.~\eqref{eq:compact_Hamiltonian_tot}, and decomposed into a near-resonant component (green solid line) and an off-resonant component (black solid lines), as given by Eq.~\eqref{eq:Eg:res}. As a guide to the eye, we also show the time evolution of the stored energy in the QB system when the terms $P_n^{z}(\phi)$ are included, again split into a near-resonant contribution (blue solid lines) and an off-resonant contribution (red solid lines), already presented in Fig.~\ref{fig:QB_full_R_NR_componets}. Panels \subref{fig:Energy_full_R_NR_vs_no_Pz_tmax_500g_0.100_phi_0.53} and \subref{fig:Energy_full_R_NR_vs_no_Pz_tmax_500g_0.300_phi_0.53} correspond to $\phi/\pi=0.53$, where the set of near-resonant TLSs is  $\mathcal{S} \equiv \{\nu = \pm 4\}$ for $g=0.1$ and $\mathcal{S} \equiv \{\nu = \pm3, \pm 4\}$ for $g=0.3$. Instead, panels \subref{fig:Energy_full_R_NR_vs_no_Pz_tmax_500g_0.100_phi_0.76} and \subref{fig:Energy_full_R_NR_vs_no_Pz_tmax_500g_0.300_phi_0.76} show the case $\phi/\pi=0.76$, for which the near-resonant TLSs set is $\mathcal{S} \equiv \{\nu = \pm 1,\pm 2\}$. Panels \subref{fig:Energy_full_R_NR_vs_no_Pz_tmax_500g_0.100_phi_0.53} and \subref{fig:Energy_full_R_NR_vs_no_Pz_tmax_500g_0.100_phi_0.76} correspond to coupling strength $g=0.1$, while \subref{fig:Energy_full_R_NR_vs_no_Pz_tmax_500g_0.300_phi_0.53} and \subref{fig:Energy_full_R_NR_vs_no_Pz_tmax_500g_0.300_phi_0.76} refer to $g=0.3$. In all panels, time is expressed in units of $\hbar/\deltaO$. The remaining parameters are $\omegar=0.75\deltaO$, $\mu_0 = 6.5\hvf/L$, and $W = 2 L$.}
  \label{fig:QB_full_vs_no_Pz_R_NR_componets}
\end{figure*}

In this Appendix, we provide a further analysis of the role played by the contribution of the longitudinal terms $P_n^z(\phi)$, which enter as first-order terms in $g$ in Eq.~\eqref{eq:compact_Hamiltonian_tot}.
In particular, Fig.~\ref{fig:QB_full_vs_no_Pz_R_NR_componets} combines what is shown in Fig.~\ref{fig:QB_full_vs_no Pz_tests} and Fig.~\ref{fig:QB_full_R_NR_componets} of the main text.
Here, the energy stored in the QB, $E^{(g)}(t)$ (in units of $N\omegar$), calculated by omitting the terms $P_n^{z}(\phi)$ in Eq.~\eqref{eq:compact_Hamiltonian_tot}, is split into a near-resonant (green solid line) and an off-resonant part (black solid lines), as specified in Eqs.~\eqref{eq:Eg:res}.
For comparison, we also show the time evolution of the stored energy in the QB system when the terms $P_n^{z}(\phi)$ are taken into account, again separated into a near-resonant contribution (blue solid lines) and an off-resonant contribution (red solid lines), as already shown in Fig.~\ref{fig:QB_full_R_NR_componets}.

In each panel, we distinguish between the contributions due to TLSs that belong to the set $\mathcal{S}$ (blue and green lines) and those that do not (red and black lines) according to the criteria introduced in the main text.
For the case $\phi/\pi=0.53$, close to the two-photon resonance, shown in Figs.~\ref{fig:Energy_full_R_NR_vs_no_Pz_tmax_500g_0.100_phi_0.53} and \subref{fig:Energy_full_R_NR_vs_no_Pz_tmax_500g_0.300_phi_0.53}, this set is defined as  $\mathcal{S} \equiv \{\nu = \pm 4\}$  for $g=0.1$ and $\mathcal{S} \equiv \{\nu = \pm 3,\pm 4\}$ for $g=0.3$ (see Fig.~\ref{fig:QB_energies} for comparison). Conversely, for $\phi/\pi=0.76$, close to the single-photon resonance, shown in Figs.~\ref{fig:Energy_full_R_NR_vs_no_Pz_tmax_500g_0.100_phi_0.76} and \subref{fig:Energy_full_R_NR_vs_no_Pz_tmax_500g_0.300_phi_0.76}, the corresponding set is $\mathcal{S} \equiv \{\nu = \pm 1,\pm 2\}$.

Specifically, Fig.~\ref{fig:Energy_full_R_NR_vs_no_Pz_tmax_500g_0.100_phi_0.53} shows the case $g=0.1$ with the phase difference set at $\phi/\pi=0.53$. As discussed in the main text, under these conditions, the charging process is expected to be dominated primarily by two-photon resonant processes.
From the comparison between the curves in Fig.~\ref{fig:Energy_full_R_NR_vs_no_Pz_tmax_500g_0.100_phi_0.53} it is evident that the terms $P_n^z(\phi)$ enhance the energy exclusively close to resonance.
Moreover, in Fig.~\ref{fig:Energy_full_R_NR_vs_no_Pz_tmax_500g_0.300_phi_0.53}, we keep the same phase difference but increase the coupling constant to $g=0.3$. We again find that the terms $P_n^z(\phi)$ act as supporting contributions that specifically affect the TLSs belonging to $\mathcal{S}$. For both values of the coupling constant, the contributions for TLSs outside the set $\mathcal{S}$ are negligible.

Figs.~\ref{fig:Energy_full_R_NR_vs_no_Pz_tmax_500g_0.100_phi_0.76} and  \subref{fig:Energy_full_R_NR_vs_no_Pz_tmax_500g_0.300_phi_0.76} refer to the single-photon resonance condition ($\phi/\pi=0.76$).
In particular, Fig.~\ref{fig:Energy_full_R_NR_vs_no_Pz_tmax_500g_0.100_phi_0.76} shows the case with $g=0.1$.
According to the perturbative analysis presented in the main text (see Eq.~\eqref{eq:Hamiltonian_int_picture} and Eq.~\eqref{eq:State_pert_evolution}), the charging dynamics is therefore expected to be predominantly governed by single-photon resonant processes. However, for $\phi/\pi = 0.76$, contributions from states outside $\mathcal{S}$ (red solid line) produce a small but noticeable contribution. This is because their effective transverse coupling $g|P_n^x(\phi)|$ is larger than that of ABSs within $\mathcal{S}$, which is, in turn, related to their transmission probabilities.

In this regime, we recall that the terms $P_n^z(\phi)$ act as a source of disturbance affecting the energy contribution of the resonant TLSs (solid blue line) belonging to $\mathcal{S}$. In fact, when the terms $P_n^z(\phi)$ are removed from the Hamiltonian, the stored energy increases (solid green line).
Keeping the same phase difference but increasing the coupling constant to
$g=0.3$, Fig.~\ref{fig:Energy_full_R_NR_vs_no_Pz_tmax_500g_0.300_phi_0.76} shows that the influence of the terms $P_n^z(\phi)$ results in a slightly enhanced overall contribution. This behavior can be interpreted as the outcome of a compensation between two competing effects. The contribution from two-photon off-resonant contributions increases, primarily due to the larger coupling constant, as indicated by the change from the black solid line to the red solid line. However, similar to what is illustrated in Fig.~\ref{fig:Energy_full_R_NR_vs_no_Pz_tmax_500g_0.100_phi_0.76}, the terms $P_n^z(\phi)$ also lead to a suppression of the contribution associated with the single-photon resonant contributions. This is demonstrated by the difference between the green solid line and the blue solid line.

\section{Near-resonant subset $\cal S$}\label{app:B}

\begin{figure*}[t]
	\centering
	\begin{subfigure}[t]{0.485\textwidth}
		\caption{}
		\includegraphics[width=\textwidth]{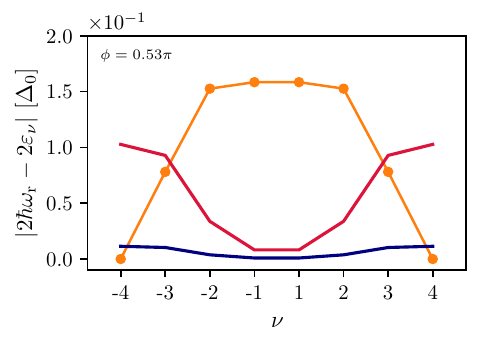}
		\label{fig:rensonant_set_2ph_vs_g}
	\end{subfigure}
	\hfill
	\begin{subfigure}[t]{0.485\textwidth}
		\caption{}
		\includegraphics[width=\textwidth]{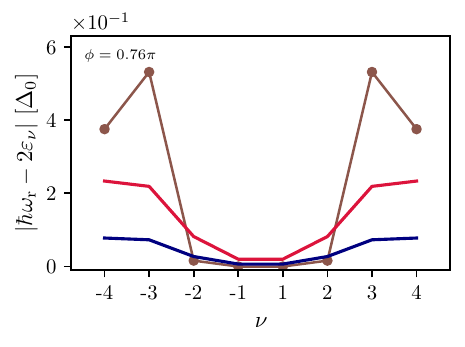}
		\label{fig:rensonant_set_1ph_vs_g}
	\end{subfigure}

	\caption{Detuning of each TLS for the near two-photon resonance case $\phi=0.53\pi$ \subref{fig:rensonant_set_2ph_vs_g} and for the near single-photon resonance case $\phi=0.76\pi$ \subref{fig:rensonant_set_1ph_vs_g} as a function of the index $\nu$, as in Fig.~\ref{fig:QB_energies}. In panel \subref{fig:rensonant_set_2ph_vs_g} they are compared with the relevant matrix element of the effective interaction at second-order in Eq.~\eqref{eq:app_two_photon_matrix_elem}, while in panel \subref{fig:rensonant_set_1ph_vs_g} they are compared with the relevant matrix element of the effective interaction at first-order in Eq.~\eqref{eq:app_single_photon_matrix_elem}.
    The symbols connected by brown/orange lines represent the TLS detunings, while the blue and red solid lines represent the absolute values of the corresponding interaction matrix elements for $g=0.1$ and $g=0.3$, respectively.
	The remaining parameters are $\omegar=0.75\deltaO$, $\mu_0 = 6.5\hvf/L$, and $W = 2 L$.}
	\label{fig:rensonant_set_criteria}
\end{figure*}

This Appendix provides a microscopic criterion for defining the subset $\cal S$ of TLSs that are resonant or near-resonant with the resonator mode. Specifically, we compare the detuning of each TLS from the relevant single- or two-photon resonance with the characteristic energy scale set by the corresponding interaction matrix element.

We recall that the initial state of the global system is
\begin{equation}
	\ket{\psi(0)} = \ket{N}\left( \mathop{\otimes}\limits_{n=1}^{N}\ket{\downarrow}_{n}\right)~.
\end{equation}

For $\phi = 0.53\pi$, corresponding to the representative case of a two-photon resonance charging process, the charging dynamics is dominated by second-order processes in which two photons are annihilated in the resonator while a single excitation is created in the TLS ensemble. Focusing on a given TLS labeled by $n$, we consider the two-photon transition
\begin{equation}
	\ket{N}\ket{\downarrow}_{n}\rightarrow \ket{N-2}\ket{\uparrow}_{n}~.
	\label{eq:app_two_photon_transition}
\end{equation}

Within a perturbative approach, according to Eq.~\eqref{eq:State_pert_evolution}, the relevant contribution to the transition amplitude is provided by the matrix element of the effective second-order operator
\begin{equation}
	\tilde{\mathcal V}_2(t) = \tilde{V}_2(t) + \tilde{V}_1(t)\int_0^{t} \frac{d t_2}{i \hbar} \tilde{V}_1(t_2)~,
	\label{eq:app_second_order_operator}
\end{equation}
between the two states appearing in Eq.~\eqref{eq:app_two_photon_transition}. In evaluating this quantity it is convenient to isolate in $\tilde{\mathcal V}_2(t)$  the terms proportional to  $\A^2 \sigmap_n$ , which are the only ones that contribute to the process in Eq.~\eqref{eq:app_two_photon_transition}. Close to the two-photon resonance condition, $2\omegar \approx 2\varepsilon_n(\phi)$, one finds
\begin{equation}
	\bra{N-2}\bra{\uparrow}_{n}\tilde{\mathcal V}_2(t) \ket{N}\ket{\downarrow}_{n}\approx \frac{g^2}{2}\left(D^x_{n}(\phi)-\frac{4P^z_{n}(\phi)P^x_{n}(\phi)}{\omegar} \right) \sqrt{N(N-1)}~,
	\label{eq:app_two_photon_matrix_elem}
\end{equation}
where the term proportional to $P_n^z(\phi)P_n^x(\phi)$ arises as a second-order correction stemming from $\tilde{V}_1(t)$. On this basis, the $n$-th TLS can be regarded as being near two-photon resonance with the resonator if its detuning satisfies the following condition
\begin{equation}
	\abs{2\omegar - 2\varepsilon_n(\phi)}\leq \frac{g^2}{2}\left|D_{n}^x(\phi) - \frac{4P_{n}^z(\phi)P_{n}^x(\phi)}{\omegar}\right|\sqrt{N(N-1)}~.
	\label{eq:app_two_photon_criterion}
\end{equation}
This inequality states that the detuning must be smaller than the characteristic energy scale set by the effective second-order interaction matrix element connecting the initial and final states.

For $\phi = 0.76\pi$, which corresponds to the representative case of a single-photon resonance charging process, the charging dynamics is instead dominated by first-order processes in which a single photon is annihilated in the resonator while an excitation is created in the TLS ensemble, namely
\begin{equation}
	\ket{N}\ket{\downarrow}_{n} \rightarrow \ket{N-1}\ket{\uparrow}_{n}~.
	\label{eq:app_single_photon_transition}
\end{equation}

In this case, the relevant energy scale entering the transition amplitude is the matrix element of the first-order interaction operator $\tilde{V}_1(t)$, defined in Eq.~\eqref{eq:Hamiltonian_int_picture}, between the initial and final state. Close to the single-photon resonance, where $\omegar \approx 2\varepsilon_n(\phi)$, one obtains
\begin{equation}
	\bra{N-1}\bra{\uparrow}_{n}\tilde{V}_1(t) \ket{N}\ket{\downarrow}_{n} \approx g P^x_{n}(\phi)\sqrt{N}~.
	\label{eq:app_single_photon_matrix_elem}
\end{equation}
Therefore, the $n$-th TLS can be regarded as near single-photon resonance with the resonator provided that
\begin{equation}
	\abs{\omegar - 2\varepsilon_n(\phi)} \leq g |P^x_{n}(\phi)|\sqrt{N}~.
	\label{eq:app_single_photon_criterion}
\end{equation}
Again, this condition requires that the detuning be smaller than the relevant first-order interaction matrix element.

The condition specified in Eq.~\eqref{eq:app_two_photon_criterion} (Eq.~\eqref{eq:app_single_photon_criterion}) formalizes the requirement that the TLS lie in the vicinity of the two-photon (single-photon) resonance with the resonator mode. Specifically, it constrains the TLS detuning from $2\omegar$ ($\omegar$) to be no larger than the characteristic energy scale defined by the corresponding relevant matrix element.
When these inequalities are satisfied, the associated TLSs undergo appreciable two-photon (single-photon) Rabi oscillations with the resonator mode, leading to a coherent exchange of energy between the two subsystems. Those TLSs therefore constitute the subset $\mathcal S$ introduced in Sec.~\ref{sec:results} of the main text. In contrast, TLSs that do not satisfy the above conditions have strongly suppressed transition rates and contribute only via virtual processes, which manifest as dispersive shifts of the resonator frequency and effective TLS-TLS interactions, as discussed in App.~\ref{app:A}.

In Fig.~\ref{fig:rensonant_set_criteria}, we compare, for both resonance conditions, the detuning of each TLS with the value of the corresponding matrix element. Specifically, Fig.~\ref{fig:rensonant_set_2ph_vs_g} shows the two-photon case at $\phi=0.53\pi$ while Fig.~\ref{fig:rensonant_set_1ph_vs_g}  refers to single-photon case where $\phi=0.76\pi$. In both situations, the matrix elements in Eqs.~\eqref{eq:app_two_photon_matrix_elem} and \eqref{eq:app_single_photon_matrix_elem} are evaluated for $g=0.1$ (blue solid line) and $g=0.3$ (red solid line). The TLSs satisfying Eq.~\eqref{eq:app_two_photon_criterion} are those whose detuning lies below the blue and red lines in Fig.~\ref{fig:rensonant_set_2ph_vs_g}.
Here, the relevant states are indexed by $\nu=\pm4$ for $g=0.1$ and by $\nu=\pm3,\pm4$ for $g=0.3$; collectively, these indices define the subset $\mathcal S$.
In the single-photon regime, the TLSs satisfying Eq.~\eqref{eq:app_single_photon_criterion} correspond to $\nu = \pm1,\pm2$ for both values of $g$.

\end{appendices}

\appendix


\section*{Acknowledgments}

G.G., F.M.D.P., and D.F. acknowledge support from the project PRIN 2022 - 2022XK5CPX (PE3) SoS-QuBa - "Solid State Quantum Batteries: Characterization and Optimization" funded within the programme "PNRR Missione 4 - Componente 2 - Investimento 1.1 Fondo per il Programma Nazionale di Ricerca e Progetti di Rilevante Interesse Nazionale (PRIN)", funded by the European Union - Next Generation EU".
V.V.  and E.P. are thankful for the PNRR MUR project PE0000023-NQSTI. E.P. acknowledges support from COST Action CA21144 superqumap, and Universit\`a degli Studi di Catania, project TCMQI PIACERI 2024/2026.

The data that support the findings of this study are openly available at the following Ref.~\cite{Varrica2026Dataset}.






\bibliographystyle{unsrt}
\bibliography{biblio_vc}

@article{beenakker1991,
  title = {{Josephson current through a superconducting quantum point contact shorter than the coherence length}},
  author = {Beenakker, C. W. J. and van Houten, H.},
  journal = {Phys. Rev. Lett.},
  volume = {66},
  issue = {23},
  pages = {3056--3059},
  numpages = {0},
  year = {1991},
  month = {Jun},
  publisher = {American Physical Society},
  doi = {10.1103/PhysRevLett.66.3056},
  url = {https://link.aps.org/doi/10.1103/PhysRevLett.66.3056}
}

@article{titovPRB2006,
  title = {{Josephson effect in ballistic graphene}},
  author = {Titov, M. and Beenakker, C. W. J.},
  journal = {Phys. Rev. B},
  volume = {74},
  issue = {4},
  pages = {041401(R)},
  numpages = {4},
  year = {2006},
  month = {Jul},
  publisher = {American Physical Society},
  doi = {10.1103/PhysRevB.74.041401},
  url = {https://link.aps.org/doi/10.1103/PhysRevB.74.041401}
}

@article{devoret2017,
  title = {{Introduction to quantum electromagnetic circuits}},
  author={Vool, Uri and Devoret, Michel},
  journal={International Journal of Circuit Theory and Applications},
  volume={45},
  url = {http://dx.doi.org/10.1002/cta.2359},
  number={7},
  pages={897--934},
  year={2017},
  publisher={Wiley Online Library}
}

@article{janvier2015,
  title = {{Coherent manipulation of Andreev states in superconducting atomic contacts}},
  volume = {349},
  ISSN = {1095-9203},
  url = {http://dx.doi.org/10.1126/science.aab2179},
  number = {6253},
  journal = {Science},
  publisher = {American Association for the Advancement of Science (AAAS)},
  author = {Janvier,  C. and Tosi,  L. and Bretheau,  L. and Girit,  undefined. \"{O}. and Stern,  M. and Bertet,  P. and Joyez,  P. and Vion,  D. and Esteve,  D. and Goffman,  M. F. and Pothier,  H. and Urbina,  C.},
  year = {2015},
  month = sep,
  pages = {1199–1202}
}

@article{wang2019,
  title = {{Coherent control of a hybrid superconducting circuit made with graphene-based van der {W}}aals heterostructures},
  volume = {14},
  url = {http://dx.doi.org/10.1038/s41565-018-0329-2},
  ISSN = {1748-3395},
  number = {2},
  journal = {Nature Nanotechnology},
  publisher = {Springer Science and Business Media LLC},
  author = {Wang,  Joel I-Jan and Rodan-Legrain,  Daniel and Bretheau,  Landry and Campbell,  Daniel L. and Kannan,  Bharath and Kim,  David and Kjaergaard,  Morten and Krantz,  Philip and Samach,  Gabriel O. and Yan,  Fei and Yoder,  Jonilyn L. and Watanabe,  Kenji and Taniguchi,  Takashi and Orlando,  Terry P. and Gustavsson,  Simon and Jarillo-Herrero,  Pablo and Oliver,  William D.},
  year = {2018},
  month = dec,
  pages = {120-125}
}

@article{schmidt2018,
  title = {{A ballistic graphene superconducting microwave circuit}},
  author={Schmidt, Felix E. and Jenkins, Mark D. and Watanabe, Kenji and Taniguchi, Takashi and Steele, Gary A.},
  journal={Nature communications},
  volume  = {9},
  number  = {1},
  pages   = {4069},
  year    = {2018},
  doi     = {10.1038/s41467-018-06595-2},
  url     = {https://doi.org/10.1038/s41467-018-06595-2}
}

@article{park2020,
  title = {{From Adiabatic to Dispersive Readout of Quantum Circuits}},
  author = {Park, Sunghun and Metzger, C. and Tosi, L. and Goffman, M. F. and Urbina, C. and Pothier, H. and Yeyati, A. Levy},
  journal = {Phys. Rev. Lett.},
  volume = {125},
    url = {http://dx.doi.org/10.1103/PhysRevLett.125.077701},
  issue = {7},
  pages = {077701},
  numpages = {5},
  year = {2020},
  month = {Aug},
  publisher = {American Physical Society},
}

@article{metzger2021,
  title = {{Circuit-{QED}} with phase-biased {J}osephson weak links},
  author = {Metzger, C. and Park, Sunghun and Tosi, L. and Janvier, C. and Reynoso, A. A. and Goffman, M. F. and Urbina, C. and Levy Yeyati, A. and Pothier, H.},
  journal = {Phys. Rev. Research},
  volume = {3},
  url = {http://dx.doi.org/10.1103/PhysRevResearch.3.013036},
  issue = {1},
  pages = {013036},
  numpages = {20},
  year = {2021},
  month = {Jan},
  publisher = {American Physical Society},
}

@article{zazunov2005,
  title = {{Dynamics and phonon-induced decoherence of Andreev level qubit}},
  author = {Zazunov, A. and Shumeiko, V. S. and Wendin, G. and Bratus', E. N.},
  journal = {Phys. Rev. B},
  volume = {71},
  issue = {21},
  pages = {214505},
  numpages = {16},
  year = {2005},
  month = {Jun},
  publisher = {American Physical Society},
  doi = {10.1103/PhysRevB.71.214505},
  url = {https://link.aps.org/doi/10.1103/PhysRevB.71.214505}
}

@article{park2018,
  title = {{Short ballistic {J}}osephson coupling in planar graphene junctions with inhomogeneous carrier doping},
  author={Park, Jinho and Lee, Jae Hyeong and Lee, Gil-Ho and Takane, Yositake and Imura, Ken-Ichiro and Taniguchi, Takashi and Watanabe, Kenji and Lee, Hu-Jong},
  journal={Physical Review Letters},
  volume={120},
  url = {http://dx.doi.org/10.1103/PhysRevLett.120.077701},
  number={7},
  pages={077701},
  year={2018},
  publisher={APS}
}

@article{kjaergaard2016,
	title = {Quantized conductance doubling and hard gap in a two-dimensional semiconductor–superconductor heterostructure},
	volume = {7},
	pages = {12841},
	ISSN = {2041-1723},
	url = {http://dx.doi.org/10.1038/ncomms12841},
	DOI = {10.1038/ncomms12841},
	number = {1},
	journal = {Nature Communications},
	publisher = {Springer Science and Business Media LLC},
	author = {Kjaergaard,  M. and Nichele,  F. and Suominen,  H. J. and Nowak,  M. P. and Wimmer,  M. and Akhmerov,  A. R. and Folk,  J. A. and Flensberg,  K. and Shabani,  J. and Palmstrøm,  C. J. and Marcus,  C. M.},
	year = {2016},
	month = sep 
}

@article{kjaergaard2017hetero,
  title = {{Transparent Semiconductor-Superconductor Interface and Induced Gap in an Epitaxial Heterostructure {J}}osephson Junction},
  author = {Kjaergaard, M. and Suominen, H. J. and Nowak, M. P. and Akhmerov, A. R. and Shabani, J. and Palmstr\o{}m, C. J. and Nichele, F. and Marcus, C. M.},
  journal = {Phys. Rev. Appl.},
  volume = {7},
  url = {http://dx.doi.org/10.1103/PhysRevApplied.7.034029},
  issue = {3},
  pages = {034029},
  numpages = {9},
  year = {2017},
  month = {Mar}, 
  publisher = {American Physical Society}
}

@article{kroll2018,
  title = {{Magnetic field compatible circuit quantum electrodynamics with graphene {J}}osephson junctions},
  author={Kroll, JG and Uilhoorn, W and van der Enden, KL and de Jong, D and Watanabe, K and Taniguchi, T and Goswami, S and Cassidy, MC and Kouwenhoven, Leo P},
  journal={Nature communications},
  volume  = {9},
  number  = {1},
  pages   = {4615},
  year    = {2018},
  doi     = {10.1038/s41467-018-07124-x},
  url     = {https://doi.org/10.1038/s41467-018-07124-x}
}

@article{mayer2019,
    author = {Mayer, William and Yuan, Joseph and Wickramasinghe, Kaushini S. and Nguyen, Tri and Dartiailh, Matthieu C. and Shabani, Javad},
    title = {Superconducting proximity effect in epitaxial Al-InAs heterostructures},
    journal = {Applied Physics Letters},
    volume = {114},
    number = {10},
    pages = {103104},
    year = {2019},
    month = {03},
    issn = {0003-6951},
    doi = {10.1063/1.5067363},
    url = {https://doi.org/10.1063/1.5067363},
    eprint = {https://pubs.aip.org/aip/apl/article-pdf/doi/10.1063/1.5067363/19772789/103104_1_online.pdf},
}

@article{goffman2017,
doi = {10.1088/1367-2630/aa7641},
url = {https://doi.org/10.1088/1367-2630/aa7641},
year = {2017},
month = {sep},
publisher = {IOP Publishing},
volume = {19},
number = {9},
pages = {092002},
author = {Goffman, M F and Urbina, C and Pothier, H and Nygård, J and Marcus, C M and Krogstrup, P},
title = {Conduction channels of an InAs-Al nanowire Josephson weak link},
journal = {New Journal of Physics}
}

@article{andolina2018,
  title = {{Charger-mediated energy transfer in exactly solvable models for quantum batteries}},
  author = {Andolina, Gian Marcello and Farina, Donato and Mari, Andrea and Pellegrini, Vittorio and Giovannetti, Vittorio and Polini, Marco},
  journal = {Phys. Rev. B},
  volume = {98},
  issue = {20},
  pages = {205423},
  numpages = {11},
  year = {2018},
  month = {Nov},
  publisher = {American Physical Society},
  doi = {10.1103/PhysRevB.98.205423},
  url = {https://link.aps.org/doi/10.1103/PhysRevB.98.205423}
}

@article{andolina2019,
  title = {{Cavity quantum electrodynamics of strongly correlated electron systems: A no-go theorem for photon condensation}},
  author = {Andolina, G. M. and Pellegrino, F. M. D. and Giovannetti, V. and MacDonald, A. H. and Polini, M.},
  journal = {Phys. Rev. B},
  volume = {100},
  issue = {12},
  pages = {121109(R)},
  numpages = {6},
  year = {2019},
  month = {Sep},
  publisher = {American Physical Society},
  doi = {10.1103/PhysRevB.100.121109},
  url = {https://link.aps.org/doi/10.1103/PhysRevB.100.121109}
}

@article{nichele2020relating,
  title = {{Relating {A}}ndreev bound states and supercurrents in hybrid {J}osephson junctions},
  author = {Nichele, F. and Portol\'es, E. and Fornieri, A. and Whiticar, A. M. and Drachmann, A. C. C. and Gronin, S. and Wang, T. and Gardner, G. C. and Thomas, C. and Hatke, A. T. and Manfra, M. J. and Marcus, C. M.},
  journal = {Phys. Rev. Lett.},
  volume = {124},
  issue = {22},
  pages = {226801},
  numpages = {6},
  year = {2020},
  month = {Jun},
  publisher = {American Physical Society},
  doi = {10.1103/PhysRevLett.124.226801},
  url = {https://link.aps.org/doi/10.1103/PhysRevLett.124.226801}
}

@article{pellegrino2022effect,
  title = {{Effect of dilute impurities on short graphene {J}}osephson junctions},
  volume = {5}, 
  url = {http://dx.doi.org/10.1038/s42005-022-01042-7},
  ISSN = {2399-3650},
  number = {1},
  journal = {Communications Physics},
  publisher = {Springer Science and Business Media LLC},
  author = {Pellegrino,  Francesco M. D. and Falci,  Giuseppe and Paladino,  Elisabetta},
  year = {2022},
  month = oct, 
  pages={265}
}

@article{elfeky2025, 
  title = {Microwave Andreev bound state spectroscopy in a semiconductor-based Planar Josephson junction},
  volume = {7},
  pages = {013248},
  ISSN = {2643-1564},
  url = {http://dx.doi.org/10.1103/physrevresearch.7.013248},
  number = {1},
  journal = {Physical Review Research},
  publisher = {American Physical Society (APS)},
  author = {Elfeky,  Bassel Heiba and Dindial,  Krishna and Brandão,  David S. and Pekerten,  Bariş and Lee,  Jaewoo and Strickland,  William M. and Strohbeen,  Patrick J. and Danilenko,  Alisa and Baker,  Lukas and Mikalsen,  Melissa and Schiela,  William and Liang,  Zixuan and Issokson,  Jacob and Levy,  Ido and Žutić,  Igor and Shabani,  Javad},
  year = {2025},
  month = mar 
}

@article{hays2018,
  title = {{Direct Microwave Measurement of Andreev-Bound-State Dynamics in a Semiconductor-Nanowire Josephson Junction}},
  author = {Hays, M. and de Lange, G. and Serniak, K. and van Woerkom, D. J. and Bouman, D. and Krogstrup, P. and Nyg\aa{}rd, J. and Geresdi, A. and Devoret, M. H.},
  journal = {Phys. Rev. Lett.},
  volume = {121},
  issue = {4},
  pages = {047001},
  numpages = {6},
  year = {2018},
  month = {Jul},
  publisher = {American Physical Society},
  doi = {10.1103/PhysRevLett.121.047001},
  url = {https://link.aps.org/doi/10.1103/PhysRevLett.121.047001}
}

@article{hays2020,
  title = {{Continuous monitoring of a trapped superconducting spin}},
  volume = {16},
  ISSN = {1745-2481},
  url = {http://dx.doi.org/10.1038/s41567-020-0952-3},
  DOI = {10.1038/s41567-020-0952-3},
  number = {11},
  journal = {Nature Physics},
  publisher = {Springer Science and Business Media LLC},
  author = {Hays,  M. and Fatemi,  V. and Serniak,  K. and Bouman,  D. and Diamond,  S. and de Lange,  G. and Krogstrup,  P. and Nygård,  J. and Geresdi,  A. and Devoret,  M. H.},
  year = {2020},
  month = jul,
  pages = {1103–1107}
}

@article{bargebos2022,
  title = {{Singlet-Doublet Transitions of a Quantum Dot Josephson Junction Detected in a Transmon Circuit}},
  author = {Bargerbos, Arno and Pita-Vidal, Marta and \ifmmode \check{Z}\else \v{Z}\fi{}itko, Rok and \'Avila, Jes\'us and Splitthoff, Lukas J. and Gr\"unhaupt, Lukas and Wesdorp, Jaap J. and Andersen, Christian K. and Liu, Yu and Kouwenhoven, Leo P. and Aguado, Ram\'on and Kou, Angela and van Heck, Bernard},
  journal = {PRX Quantum},
  volume = {3},
  issue = {3},
  pages = {030311},
  numpages = {14},
  year = {2022},
  month = {Jul},
  publisher = {American Physical Society},
  doi = {10.1103/PRXQuantum.3.030311},
  url = {https://link.aps.org/doi/10.1103/PRXQuantum.3.030311}
}

@article{pitavidal2023,
title = {{Direct manipulation of a superconducting spin qubit strongly coupled to a transmon qubit}},
volume = {19},
ISSN = {1745-2481},
url = {http://dx.doi.org/10.1038/s41567-023-02071-x},
DOI = {10.1038/s41567-023-02071-x},
number = {8},
journal = {Nature Physics},
publisher = {Springer Science and Business Media LLC},
author = {Pita-Vidal,  Marta and Bargerbos,  Arno and Žitko,  Rok and Splitthoff,  Lukas J. and Gr\"{u}nhaupt,  Lukas and Wesdorp,  Jaap J. and Liu,  Yu and Kouwenhoven,  Leo P. and Aguado,  Ramón and van Heck,  Bernard and Kou,  Angela and Andersen,  Christian Kraglund},
year = {2023},
month = may,
pages = {1110–1115}
}

@article{hinderling2023,
  title = {Flip-Chip-Based Microwave Spectroscopy of Andreev Bound States in a Planar Josephson Junction},
  volume = {19},
  pages = {054026},
  ISSN = {2331-7019},
  url = {http://dx.doi.org/10.1103/PhysRevApplied.19.054026},
  number = {5},
  journal = {Physical Review Applied},
  publisher = {American Physical Society (APS)},
  author = {Hinderling,  M. and Sabonis,  D. and Paredes,  S. and Haxell,  D.Z. and Coraiola,  M. and ten Kate,  S.C. and Cheah,  E. and Krizek,  F. and Schott,  R. and Wegscheider,  W. and Nichele,  F.},
  year = {2023},
  month = may 
}

@article{haller2022,
  title = {{Phase-dependent microwave response of a graphene Josephson junction}},
  author = {Haller, R. and F\"ul\"op, G. and Indolese, D. and Ridderbos, J. and Kraft, R. and Cheung, L. Y. and Ungerer, J. H. and Watanabe, K. and Taniguchi, T. and Beckmann, D. and Danneau, R. and Virtanen, P. and Sch\"onenberger, C.},
  journal = {Phys. Rev. Res.},
  volume = {4},
  issue = {1},
  pages = {013198},
  numpages = {11},
  year = {2022},
  month = {Mar},
  publisher = {American Physical Society},
  doi = {10.1103/PhysRevResearch.4.013198},
  url = {https://link.aps.org/doi/10.1103/PhysRevResearch.4.013198}
}

@article{iorio2023,
  title = {{Half-integer {S}}hapiro steps in highly transmissive {I}n{S}b nanoflag {J}osephson junctions},
  author = {Iorio, A. and Crippa, A. and Turini, B. and Salimian, S. and Carrega, M. and Chirolli, L. and Zannier, V. and Sorba, L. and Strambini, E. and Giazotto, F. and Heun, S.},
  journal = {Phys. Rev. Res.},
  volume = {5},
  issue = {3},
  pages = {033015},
  numpages = {15},
  year = {2023},
  month = {Jul},
  publisher = {American Physical Society},
  doi = {10.1103/PhysRevResearch.5.033015},
  url = {https://link.aps.org/doi/10.1103/PhysRevResearch.5.033015}
}

@article{turini2024,
  title = {{Side‐Gate Modulation of Supercurrent in {I}}n{S}b {N}anoflag‐{B}ased {J}osephson {J}unctions},
  ISSN = {1521-3951},
  url = {http://dx.doi.org/10.1002/pssb.202400534},
  journal = {physica status solidi (b)},
  publisher = {Wiley},
  author = {Turini,  Bianca and Salimian,  Sedighe and Carrega,  Matteo and Paolucci,  Federico and Zannier,  Valentina and Sorba,  Lucia and Heun,  Stefan},
  year = {2024},
  month = dec 
}

@article{monroe2024,
  title = {{Phase jumps in Josephson junctions with time-dependent spin–orbit coupling}},
  volume = {125},
  ISSN = {1077-3118},
  url = {http://dx.doi.org/10.1063/5.0211562},
  DOI = {10.1063/5.0211562},
  number = {1},
  journal = {Applied Physics Letters},
  publisher = {AIP Publishing},
  author = {Monroe, David and Shen, Chenghao and Tringali, Dario and Alidoust, Mohammad and Zhou, Tong and Žutić, Igor},
  year = {2024},
  month = jul 
}

@article{strickland2024,
  title = {{Characterizing losses in InAs two-dimensional electron gas-based gatemon qubits}},
  author = {Strickland, William M. and Baker, Lukas J. and Lee, Jaewoo and Dindial, Krishna and Elfeky, Bassel Heiba and Stroh, Patrick J. and Hatefipour, Mehdi and Yu, Peng and Levy, Ido and Issokson, Jacob and Manucharyan, Vladimir E. and Shabani, Javad},
  journal = {Phys. Rev. Res.},
  volume = {6},
  url = {http://dx.doi.org/10.1103/PhysRevResearch.6.023094},
  issue = {2},
  pages = {023094},
  numpages = {12},
  year = {2024},
  month = {Apr},
  publisher = {American Physical Society}
}

@article{marcus2023planar,
  title = {{Signatures of a topological phase transition in a planar Josephson junction}},
  author = {Banerjee, A. and Lesser, O. and Rahman, M. A. and Wang, H.-R. and Li, M.-R. and Kringh\o{}j, A. and Whiticar, A. M. and Drachmann, A. C. C. and Thomas, C. and Wang, T. and Manfra, M. J. and Berg, E. and Oreg, Y. and Stern, Ady and Marcus, C. M.},
  journal = {Phys. Rev. B},
  volume = {107},
  url = {http://dx.doi.org/10.1103/PhysRevB.107.245304},
  issue = {24},
  pages = {245304},
  numpages = {12},
  year = {2023},
  month = {Jun},
  publisher = {American Physical Society}
}

@inbook{Beenakker1992,
  title = {{Three “Universal” Mesoscopic Josephson Effects}},
  ISBN = {9783642848186},
  ISSN = {0171-1873},
  url = {http://dx.doi.org/10.1007/978-3-642-84818-6_22},
  DOI = {10.1007/978-3-642-84818-6_22},
  booktitle = {{Transport Phenomena in Mesoscopic Systems}},
  publisher = {Springer Berlin Heidelberg},
  author = {Beenakker,  C. W. J.},
  year = {1992},
  pages = {235–253}
}

@phdthesis{bretheau2013,
  title = {{{Localized Excitations in Superconducting Atomic Contacts: probing the {A}}ndreev doublet}},
  AUTHOR = {Bretheau, Landry},
  SCHOOL = {{Ecole Polytechnique}},
  YEAR = {2013},
  URL = {https://pastel.hal.science/pastel-00862029},
  MONTH = Feb,
  TYPE = {PhD Thesis},
  HAL_ID = {pastel-00862029},
  HAL_VERSION = {v1},
}

@article{bryon2023,
  title = {{Time-Dependent Magnetic Flux in Devices for Circuit Quantum Electrodynamics}},
  author = {Bryon, Jacob and Weiss, D.K. and You, Xinyuan and Sussman, Sara and Croot, Xanthe and Huang, Ziwen and Koch, Jens and Houck, Andrew A.},
  journal = {Phys. Rev. Appl.},
  volume = {19},
  issue = {3},
  pages = {034031},
  numpages = {9},
  year = {2023},
  month = {Mar},
  publisher = {American Physical Society},
  doi = {10.1103/PhysRevApplied.19.034031},
  url = {https://link.aps.org/doi/10.1103/PhysRevApplied.19.034031}
}

@article{hinderling2024,
  title = {{Direct Microwave Spectroscopy of Andreev Bound States in Planar }$\mathrm{Ge}$ {Josephson Junctions}},
  author = {Hinderling, M. and ten Kate, S. C. and Coraiola, M. and Haxell, D.Z. and Stiefel, M. and Mergenthaler, M. and Paredes, S. and Bedell, S.W. and Sabonis, D. and Nichele, F.},
  journal = {PRX Quantum},
  volume = {5},
  issue = {3},
  pages = {030357},
  numpages = {14},
  year = {2024},
  month = {Sep},
  publisher = {American Physical Society},
  doi = {10.1103/PRXQuantum.5.030357},
  url = {https://link.aps.org/doi/10.1103/PRXQuantum.5.030357}
}

@article{dicke1954,
  title = {{Coherence in Spontaneous Radiation Processes}},
  author = {Dicke, R. H.},
  journal = {Phys. Rev.},
  volume = {93},
  issue = {1},
  pages = {99--110},
  numpages = {0},
  year = {1954},
  month = {Jan},
  publisher = {American Physical Society},
  doi = {10.1103/PhysRev.93.99},
  url = {https://link.aps.org/doi/10.1103/PhysRev.93.99}
}

@article{kirton2018,
  author = {Kirton, Peter and Roses, M. M. and Keeling, Jonathan and Dalla Torre, E. G.},
  title = {{Introduction to the Dicke Model: From Equilibrium to Nonequilibrium, and Vice Versa}},
  journal = {Advanced Quantum Technologies},
  volume = {2},
  number = {1-2},
  pages = {1800043},
  year = {2019},
  url = { https://doi.org/10.1002/qute.201800043},
  doi = {10.1002/qute.201800043}
}

@article{ferraro2018high,
  title = {{High-Power Collective Charging of a Solid-State Quantum Battery}},
  author = {Ferraro, Dario and Campisi, Michele and Andolina, Gian Marcello and Pellegrini, Vittorio and Polini, Marco},
  journal = {Phys. Rev. Lett.},
  volume = {120},
  issue = {11},
  pages = {117702},
  numpages = {6},
  year = {2018},
  month = {Mar},
  publisher = {American Physical Society},
  doi = {10.1103/PhysRevLett.120.117702},
  url = {https://link.aps.org/doi/10.1103/PhysRevLett.120.117702}
}

@article{andolina2019work,
  title = {{Extractable Work, the Role of Correlations, and Asymptotic Freedom in Quantum Batteries}},
  author = {Andolina, Gian Marcello and Keck, Maximilian and Mari, Andrea and Campisi, Michele and Giovannetti, Vittorio and Polini, Marco},
  journal = {Phys. Rev. Lett.},
  volume = {122},
  issue = {4},
  pages = {047702},
  numpages = {5},
  year = {2019},
  month = {Feb},
  publisher = {American Physical Society},
  doi = {10.1103/PhysRevLett.122.047702},
  url = {https://link.aps.org/doi/10.1103/PhysRevLett.122.047702}
}

@article{crescente2020ultrafast,
  title = {{Ultrafast charging in a two-photon Dicke quantum battery}},
  author = {Crescente, Alba and Carrega, Matteo and Sassetti, Maura and Ferraro, Dario},
  journal = {Phys. Rev. B},
  volume = {102},
  issue = {24},
  pages = {245407},
  numpages = {12},
  year = {2020},
  month = {Dec},
  publisher = {American Physical Society},
  doi = {10.1103/PhysRevB.102.245407},
  url = {https://link.aps.org/doi/10.1103/PhysRevB.102.245407}
}

@article{gemme2023off,
AUTHOR = {Gemme, Giulia and Andolina, Gian Marcello and Pellegrino, Francesco Maria Dimitri and Sassetti, Maura and Ferraro, Dario},
title = {{Off-Resonant Dicke Quantum Battery: Charging by Virtual Photons}},
JOURNAL = {Batteries},
VOLUME = {9},
YEAR = {2023},
NUMBER = {4},
ARTICLE-NUMBER = {197},
URL = {https://www.mdpi.com/2313-0105/9/4/197},
ISSN = {2313-0105},
DOI = {10.3390/batteries9040197}
}

@article{elghaayda2025sqb,
  title = {{Performance of a Superconducting Quantum Battery}},
  ISSN = {2511-9044},
  url = {http://dx.doi.org/10.1002/qute.202400651},
  journal = {Advanced Quantum Technologies},
  publisher = {Wiley},
  author = {Elghaayda,  Samira and Ali,  Asad and Al‐Kuwari,  Saif and Czerwinski,  Artur and Mansour,  Mostafa and Haddadi,  Saeed},
  year = {2025},
  month = mar 
}

@article{razzoli2025cyclic,
doi = {10.1088/2058-9565/ad9ed4},
url = {https://doi.org/10.1088/2058-9565/ad9ed4},
year = {2025},
month = {jan},
publisher = {IOP Publishing},
volume = {10},
number = {1},
pages = {015064},
author = {Razzoli, Luca and Gemme, Giulia and Khomchenko, Ilia and Sassetti, Maura and Ouerdane, Henni and Ferraro, Dario and Benenti, Giuliano},
title = {{Cyclic solid-state quantum battery: thermodynamic characterization and quantum hardware simulation}},
journal = {Quantum Science and Technology}
}

@article{pellegrino_prb_2014,
  title = {{Theory of integer quantum Hall polaritons in graphene}},
  author = {Pellegrino, F. M. D. and Chirolli, L. and Fazio, Rosario and Giovannetti, V. and Polini, Marco},
  journal = {Phys. Rev. B},
  volume = {89},
  issue = {16},
  pages = {165406},
  numpages = {17},
  year = {2014},
  month = {Apr},
  publisher = {American Physical Society},
  doi = {10.1103/PhysRevB.89.165406},
  url = {https://link.aps.org/doi/10.1103/PhysRevB.89.165406}
}

@article{Felicetti15,
  title = {{Spectral collapse via two-phonon interactions in trapped ions}},
  author = {Felicetti, S. and Pedernales, J. S. and Egusquiza, I. L. and Romero, G. and Lamata, L. and Braak, D. and Solano, E.},
  journal = {Phys. Rev. A},
  volume = {92},
  issue = {3},
  pages = {033817},
  numpages = {7},
  year = {2015},
  month = {Sep},
  publisher = {American Physical Society},
  doi = {10.1103/PhysRevA.92.033817},
  url = {https://link.aps.org/doi/10.1103/PhysRevA.92.033817}
}

@article{Felicetti18,
  title = {{Two-photon quantum Rabi model with superconducting circuits}},
  author = {Felicetti, S. and Rossatto, D. Z. and Rico, E. and Solano, E. and Forn-D\'{\i}az, P.},
  journal = {Phys. Rev. A},
  volume = {97},
  issue = {1},
  pages = {013851},
  numpages = {11},
  year = {2018},
  month = {Jan},
  publisher = {American Physical Society},
  doi = {10.1103/PhysRevA.97.013851},
  url = {https://link.aps.org/doi/10.1103/PhysRevA.97.013851}
}

@Article{Nataf10,
author={Nataf, Pierre
and Ciuti, Cristiano},
title = {{No-go theorem for superradiant quantum phase transitions in cavity QED and counter-example in circuit QED}},
journal={Nature Communications},
year={2010},
month={Sep},
day={07},
volume={1},
number={1},
pages={72},
issn={2041-1723},
doi={10.1038/ncomms1069},
url={https://doi.org/10.1038/ncomms1069}
}

@article{sete2021,
  title = {{Floating Tunable Coupler for Scalable Quantum Computing Architectures}},
  author = {Sete, Eyob A. and Chen, Angela Q. and Manenti, Riccardo and Kulshreshtha, Shobhan and Poletto, Stefano},
  journal = {Phys. Rev. Appl.},
  volume = {15},
  issue = {6},
  pages = {064063},
  numpages = {12},
  year = {2021},
  month = {Jun},
  publisher = {American Physical Society},
  doi = {10.1103/PhysRevApplied.15.064063},
  url = {https://link.aps.org/doi/10.1103/PhysRevApplied.15.064063}
}

@article{campbell2023,
  title = {{Modular Tunable Coupler for Superconducting Circuits}},
  author = {Campbell, Daniel L. and Kamal, Archana and Ranzani, Leonardo and Senatore, Michael and LaHaye, Matthew D.},
  journal = {Phys. Rev. Appl.},
  volume = {19},
  issue = {6},
  pages = {064043},
  numpages = {17},
  year = {2023},
  month = {Jun},
  publisher = {American Physical Society},
  doi = {10.1103/PhysRevApplied.19.064043},
  url = {https://link.aps.org/doi/10.1103/PhysRevApplied.19.064043}
}

@article{heunisch2023,
  title = {{Tunable coupler to fully decouple and maximally localize superconducting qubits}},
  author = {Heunisch, Lukas and Eichler, Christopher and Hartmann, Michael J.},
  journal = {Phys. Rev. Appl.},
  volume = {20},
  issue = {6},
  pages = {064037},
  numpages = {14},
  year = {2023},
  month = {Dec},
  publisher = {American Physical Society},
  doi = {10.1103/PhysRevApplied.20.064037},
  url = {https://link.aps.org/doi/10.1103/PhysRevApplied.20.064037}
}

@article{Kurman25,
  title = {{Powering Quantum Computation with Quantum Batteries}},
  author = {Kurman, Yaniv and Hymas, Kieran and Fedorov, Arkady and Munro, William J. and Quach, James},
  journal = {Phys. Rev. X},
  volume = {16},
  issue = {1},
  pages = {011016},
  numpages = {17},
  year = {2026},
  month = {Jan},
  publisher = {American Physical Society},
  doi = {10.1103/l39v-jwwz},
  url = {https://link.aps.org/doi/10.1103/l39v-jwwz}
}

@book{sakurai_book,
      author        = "Sakurai, Jun John",
      title         = "{Modern Quantum Mechanics}",
      publisher     = "Addison-Wesley",
      address       = "Reading, MA",
      year          = "1994"
}

@book{GarciaRipoll_2022, 
    place={Cambridge}, 
    title = {{Quantum Information and Quantum Optics with Superconducting Circuits}}, publisher={Cambridge University Press}, 
    author={Garcia-Ripoll, Juan Jose}, 
    year={2022}}

@article{tsyplyatyev2009,
  title = {{Dynamics of the inhomogeneous Dicke model for a single-boson mode coupled to a bath of nonidentical spin-1/2 systems}},
  author = {Tsyplyatyev, Oleksandr and Loss, Daniel},
  journal = {Phys. Rev. A},
  volume = {80},
  issue = {2},
  pages = {023803},
  numpages = {6},
  year = {2009},
  month = {Aug},
  publisher = {American Physical Society},
  doi = {10.1103/PhysRevA.80.023803},
  url = {https://link.aps.org/doi/10.1103/PhysRevA.80.023803}
}

@article{diniz2011,
  title = {{Strongly coupling a cavity to inhomogeneous ensembles of emitters: Potential for long-lived solid-state quantum memories}},
  author = {Diniz, I. and Portolan, S. and Ferreira, R. and G\'erard, J. M. and Bertet, P. and Auff\`eves, A.},
  journal = {Phys. Rev. A},
  volume = {84},
  issue = {6},
  pages = {063810},
  numpages = {9},
  year = {2011},
  month = {Dec},
  publisher = {American Physical Society},
  doi = {10.1103/PhysRevA.84.063810},
  url = {https://link.aps.org/doi/10.1103/PhysRevA.84.063810}
}

@article{strter2012,
  title = {{Nonequilibrum dynamics in the strongly excited inhomogeneous Dicke model}},
  author = {Str\"ater, Christoph and Tsyplyatyev, Oleksandr and Faribault, Alexandre},
  journal = {Phys. Rev. B},
  volume = {86},
  issue = {19},
  pages = {195101},
  numpages = {9},
  year = {2012},
  month = {Nov},
  publisher = {American Physical Society},
  doi = {10.1103/PhysRevB.86.195101},
  url = {https://link.aps.org/doi/10.1103/PhysRevB.86.195101}
}

@misc{Hadipour2026,
      title={Cavity-Mediated Charging of a Graphene Excitonic Quantum Battery}, 
      author={Maryam Hadipour and Soroush Haseli},
      year={2026},
      eprint={2607.18489},
      archivePrefix={arXiv},
      primaryClass={quant-ph},
      url={https://arxiv.org/abs/2607.18489}, 
}

@article{Varrica2026,
  title = {Hybrid light-matter excitations and spontaneous time-reversal symmetry breaking in two-dimensional Josephson junctions},
  author = {Varrica, V. and Falci, G. and Paladino, E. and Pellegrino, F. M. D.},
  journal = {Phys. Rev. B},
  volume = {114},
  issue = {11},
  pages = {115401},
  numpages = {24},
  year = {2026},
  month = {Aug},
  publisher = {American Physical Society},
  doi = {10.1103/npcs-dl1q},
  url = {https://link.aps.org/doi/10.1103/npcs-dl1q}
}

@article{Campaioli24,
  title = {{Colloquium: Quantum batteries}},
  author = {Campaioli, Francesco and Gherardini, Stefano and Quach, James Q. and Polini, Marco and Andolina, Gian Marcello},
  journal = {Rev. Mod. Phys.},
  volume = {96},
  issue = {3},
  pages = {031001},
  numpages = {30},
  year = {2024},
  month = {Jul},
  publisher = {American Physical Society},
  doi = {10.1103/RevModPhys.96.031001},
  url = {https://link.aps.org/doi/10.1103/RevModPhys.96.031001}
}

@article{Quach23,
title = {{Quantum batteries: The future of energy storage?}},
journal = {Joule},
volume = {7},
number = {10},
pages = {2195-2200},
year = {2023},
issn = {2542-4351},
doi = {https://doi.org/10.1016/j.joule.2023.09.003},
url = {https://www.sciencedirect.com/science/article/pii/S2542435123003641},
author = {J.Q. Quach and G. Cerullo and T. Virgili}
}

@article{Camposeo25,
author = {Camposeo, Andrea and Virgili, Tersilla and Lombardi, Floriana and Cerullo, Giulio and Pisignano, Dario and Polini, Marco},
title = {{Quantum Batteries: A Materials Science Perspective}},
journal = {Advanced Materials},
volume = {37},
number = {17},
pages = {2415073},
doi = {https://doi.org/10.1002/adma.202415073},
url = {https://advanced.onlinelibrary.wiley.com/doi/abs/10.1002/adma.202415073},
year = {2025}
}

@Article{Ferraro26,
author={Ferraro, Dario
and Cavaliere, Fabio
and Genoni, Marco G.
and Benenti, Giuliano
and Sassetti, Maura},
title = {{Opportunities and challenges of quantum batteries}},
journal={Nature Reviews Physics},
year={2026},
month={Feb},
day={01},
volume={8},
number={2},
pages={115-127},
issn={2522-5820},
doi={10.1038/s42254-025-00906-5},
url={https://doi.org/10.1038/s42254-025-00906-5}
}

@misc{ezratty2025understandingquantumtechnologies2025,
      title = {{Understanding Quantum Technologies 2025}}, 
      author={Olivier Ezratty},
      year={2025},
      eprint={2111.15352},
      archivePrefix={arXiv},
      primaryClass={quant-ph},
      url={https://arxiv.org/abs/2111.15352}, 
}

@article{Auffeves22,
  title = {{Quantum Technologies Need a Quantum Energy Initiative}},
  author = {Auff\`eves, Alexia},
  journal = {PRX Quantum},
  volume = {3},
  issue = {2},
  pages = {020101},
  numpages = {12},
  year = {2022},
  month = {Jun},
  publisher = {American Physical Society},
  doi = {10.1103/PRXQuantum.3.020101},
  url = {https://link.aps.org/doi/10.1103/PRXQuantum.3.020101}
}

@article{quach2022micro,
author = {James Q. Quach  and Kirsty E. McGhee  and Lucia Ganzer  and Dominic M. Rouse  and Brendon W. Lovett  and Erik M. Gauger  and Jonathan Keeling  and Giulio Cerullo  and David G. Lidzey  and Tersilla Virgili },
title = {{Superabsorption in an organic microcavity: Toward a quantum battery}},
journal = {Science Advances},
volume = {8},
number = {2},
pages = {eabk3160},
year = {2022},
doi = {10.1126/sciadv.abk3160},
URL = {https://www.science.org/doi/abs/10.1126/sciadv.abk3160}}

@article{Rodriguez23,
  title = {{Artificial intelligence discovery of a charging protocol in a micromaser quantum battery}},
  author = {Rodr\'{\i}guez, Carla and Rosa, Dario and Olle, Jan},
  journal = {Phys. Rev. A},
  volume = {108},
  issue = {4},
  pages = {042618},
  numpages = {10},
  year = {2023},
  month = {Oct},
  publisher = {American Physical Society},
  doi = {10.1103/PhysRevA.108.042618},
  url = {https://link.aps.org/doi/10.1103/PhysRevA.108.042618}
}

@article{Erdman24,
  title = {{Reinforcement Learning Optimization of the Charging of a Dicke Quantum Battery}},
  author = {Erdman, Paolo Andrea and Andolina, Gian Marcello and Giovannetti, Vittorio and No\'e, Frank},
  journal = {Phys. Rev. Lett.},
  volume = {133},
  issue = {24},
  pages = {243602},
  numpages = {7},
  year = {2024},
  month = {Dec},
  publisher = {American Physical Society},
  doi = {10.1103/PhysRevLett.133.243602},
  url = {https://link.aps.org/doi/10.1103/PhysRevLett.133.243602}
}

@article{Tibben25,
  title = {{Extending the Self-Discharge Time of Dicke Quantum Batteries Using Molecular Triplets}},
  author = {Tibben, Daniel J. and Della Gaspera, Enrico and van Embden, Joel and Reineck, Philipp and Quach, James Q. and Campaioli, Francesco and G\'omez, Daniel E.},
  journal = {PRX Energy},
  volume = {4},
  issue = {2},
  pages = {023012},
  numpages = {10},
  year = {2025},
  month = {Jun},
  publisher = {American Physical Society},
  doi = {10.1103/bhyh-53np},
  url = {https://link.aps.org/doi/10.1103/bhyh-53np}
}

@Article{Hymas26,
author={Hymas, Kieran
and Muir, Jack B.
and Tibben, Daniel
and van Embden, Joel
and Hirai, Tadahiko
and Dunn, Christopher J.
and G{\'o}mez, Daniel E.
and Hutchison, James A.
and Smith, Trevor A.
and Quach, James Q.},
title = {{Superextensive electrical power from a quantum battery}},
journal={Light: Science {\&} Applications},
year={2026},
month={Mar},
day={13},
volume={15},
number={1},
pages={168},
issn={2047-7538},
doi={10.1038/s41377-026-02240-6},
url={https://doi.org/10.1038/s41377-026-02240-6}
}

@article{Julia20,
  title = {{Bounds on the capacity and power of quantum batteries}},
  author = {Juli\`a-Farr\'e, Sergi and Salamon, Tymoteusz and Riera, Arnau and Bera, Manabendra N. and Lewenstein, Maciej},
  journal = {Phys. Rev. Res.},
  volume = {2},
  issue = {2},
  pages = {023113},
  numpages = {16},
  year = {2020},
  month = {May},
  publisher = {American Physical Society},
  doi = {10.1103/PhysRevResearch.2.023113},
  url = {https://link.aps.org/doi/10.1103/PhysRevResearch.2.023113}
}

@article{Andolina19,
  title = {{Quantum versus classical many-body batteries}},
  author = {Andolina, Gian Marcello and Keck, Maximilian and Mari, Andrea and Giovannetti, Vittorio and Polini, Marco},
  journal = {Phys. Rev. B},
  volume = {99},
  issue = {20},
  pages = {205437},
  numpages = {7},
  year = {2019},
  month = {May},
  publisher = {American Physical Society},
  doi = {10.1103/PhysRevB.99.205437},
  url = {https://link.aps.org/doi/10.1103/PhysRevB.99.205437}
}

@article{Ferraro19,
  title = {{Quantum supercapacitors}},
  author = {Ferraro, Dario and Andolina, Gian Marcello and Campisi, Michele and Pellegrini, Vittorio and Polini, Marco},
  journal = {Phys. Rev. B},
  volume = {100},
  issue = {7},
  pages = {075433},
  numpages = {9},
  year = {2019},
  month = {Aug},
  publisher = {American Physical Society},
  doi = {10.1103/PhysRevB.100.075433},
  url = {https://link.aps.org/doi/10.1103/PhysRevB.100.075433}
}

@article{Sun25,
doi = {10.1088/1367-2630/ae2a62},
url = {https://doi.org/10.1088/1367-2630/ae2a62},
year = {2025},
month = {dec},
publisher = {IOP Publishing},
volume = {27},
number = {12},
pages = {124513},
author = {Sun, Peng-Yu and Zhou, Hang and Dou, Fu-Quan},
title = {{Cavity-Heisenberg spin-j chain quantum battery and reinforcement learning optimization}},
journal = {New Journal of Physics}
}

@article{Yang24,
  title = {{Three-level Dicke quantum battery}},
  author = {Yang, Dong-Lin and Yang, Fang-Mei and Dou, Fu-Quan},
  journal = {Phys. Rev. B},
  volume = {109},
  issue = {23},
  pages = {235432},
  numpages = {12},
  year = {2024},
  month = {Jun},
  publisher = {American Physical Society},
  doi = {10.1103/PhysRevB.109.235432},
  url = {https://link.aps.org/doi/10.1103/PhysRevB.109.235432}
}

@article{Chiribella21,
  title = {{Fundamental Energy Requirement of Reversible Quantum Operations}},
  author = {Chiribella, Giulio and Yang, Yuxiang and Renner, Renato},
  journal = {Phys. Rev. X},
  volume = {11},
  issue = {2},
  pages = {021014},
  numpages = {11},
  year = {2021},
  month = {Apr},
  publisher = {American Physical Society},
  doi = {10.1103/PhysRevX.11.021014},
  url = {https://link.aps.org/doi/10.1103/PhysRevX.11.021014}
}

@article{Alicki13,
  title = {{Entanglement boost for extractable work from ensembles of quantum batteries}},
  author = {Alicki, Robert and Fannes, Mark},
  journal = {Phys. Rev. E},
  volume = {87},
  issue = {4},
  pages = {042123},
  numpages = {4},
  year = {2013},
  month = {Apr},
  publisher = {American Physical Society},
  doi = {10.1103/PhysRevE.87.042123},
  url = {https://link.aps.org/doi/10.1103/PhysRevE.87.042123}
}

@article{Binder15,
doi = {10.1088/1367-2630/17/7/075015},
url = {https://doi.org/10.1088/1367-2630/17/7/075015},
year = {2015},
month = {jul},
publisher = {IOP Publishing},
volume = {17},
number = {7},
pages = {075015},
author = {Binder, Felix C and Vinjanampathy, Sai and Modi, Kavan and Goold, John},
title = {{Quantacell: powerful charging of quantum batteries}},
journal = {New Journal of Physics}
}

@article{Le18,
  title = {{Spin-chain model of a many-body quantum battery}},
  author = {Le, Thao P. and Levinsen, Jesper and Modi, Kavan and Parish, Meera M. and Pollock, Felix A.},
  journal = {Phys. Rev. A},
  volume = {97},
  issue = {2},
  pages = {022106},
  numpages = {9},
  year = {2018},
  month = {Feb},
  publisher = {American Physical Society},
  doi = {10.1103/PhysRevA.97.022106},
  url = {https://link.aps.org/doi/10.1103/PhysRevA.97.022106}
}

@article{Grazi24,
  title = {{Controlling Energy Storage Crossing Quantum Phase Transitions in an Integrable Spin Quantum Battery}},
  author = {Grazi, Riccardo and Sacco Shaikh, Daniel and Sassetti, Maura and Traverso Ziani, Niccol\'o and Ferraro, Dario},
  journal = {Phys. Rev. Lett.},
  volume = {133},
  issue = {19},
  pages = {197001},
  numpages = {6},
  year = {2024},
  month = {Nov},
  publisher = {American Physical Society},
  doi = {10.1103/PhysRevLett.133.197001},
  url = {https://link.aps.org/doi/10.1103/PhysRevLett.133.197001}
}

@article{Rossini19,
  title = {{Many-body localized quantum batteries}},
  author = {Rossini, Davide and Andolina, Gian Marcello and Polini, Marco},
  journal = {Phys. Rev. B},
  volume = {100},
  issue = {11},
  pages = {115142},
  numpages = {11},
  year = {2019},
  month = {Sep},
  publisher = {American Physical Society},
  doi = {10.1103/PhysRevB.100.115142},
  url = {https://link.aps.org/doi/10.1103/PhysRevB.100.115142}
}

@article{Catalano24,
  title = {{Frustrating Quantum Batteries}},
  author = {Catalano, A.G. and Giampaolo, S.M. and Morsch, O. and Giovannetti, V. and Franchini, F.},
  journal = {PRX Quantum},
  volume = {5},
  issue = {3},
  pages = {030319},
  numpages = {15},
  year = {2024},
  month = {Jul},
  publisher = {American Physical Society},
  doi = {10.1103/PRXQuantum.5.030319},
  url = {https://link.aps.org/doi/10.1103/PRXQuantum.5.030319}
}

@article{Grazi25,
title = {{Charging free fermion quantum batteries}},
journal = {Chaos, Solitons \& Fractals},
volume = {196},
pages = {116383},
year = {2025},
issn = {0960-0779},
doi = {https://doi.org/10.1016/j.chaos.2025.116383},
url = {https://www.sciencedirect.com/science/article/pii/S0960077925003960},
author = {Riccardo Grazi and Fabio Cavaliere and Maura Sassetti and Dario Ferraro and Niccolò {Traverso Ziani}}
}

@article{Cavaliere25,
  author = {Cavaliere, Fabio and Gemme, Giulia and Benenti, Giuliano and Ferraro, Dario and Sassetti, Maura},
  title = {{Dynamical blockade of a reservoir for optimal performances of a quantum battery}},
  journal = {Communications Physics},
  volume = {8},
  number = {1},
  pages = {76},
  year = {2025},
  doi = {10.1038/s42005-025-01993-7},
  url = {https://doi.org/10.1038/s42005-025-01993-7}
}

@article{Cavaliere26,
  title = {Quantum advantage bounds for a multipartite Gaussian battery},
  author = {Cavaliere, F. and Ferraro, D. and Carrega, M. and Benenti, G. and Sassetti, M.},
  journal = {Phys. Rev. Res.},
  volume = {8},
  issue = {3},
  pages = {033045},
  numpages = {16},
  year = {2026},
  month = {Jul},
  publisher = {American Physical Society},
  doi = {10.1103/r8v1-5xwf},
  url = {https://link.aps.org/doi/10.1103/r8v1-5xwf}
}

@article{Hovhannisyan20,
  title = {{Charging assisted by thermalization}},
  author = {Hovhannisyan, Karen V. and Barra, Felipe and Imparato, Alberto},
  journal = {Phys. Rev. Res.},
  volume = {2},
  issue = {3},
  pages = {033413},
  numpages = {24},
  year = {2020},
  month = {Sep},
  publisher = {American Physical Society},
  doi = {10.1103/PhysRevResearch.2.033413},
  url = {https://link.aps.org/doi/10.1103/PhysRevResearch.2.033413}
}

@article{Barra22,
doi = {10.1088/1367-2630/ac43ed},
url = {https://doi.org/10.1088/1367-2630/ac43ed},
year = {2022},
month = {jan},
publisher = {IOP Publishing},
volume = {24},
number = {1},
pages = {015003},
author = {Barra, Felipe and Hovhannisyan, Karen V and Imparato, Alberto},
title = {{Quantum batteries at the verge of a phase transition}},
journal = {New Journal of Physics}
}

@article{Carrasco22,
  title = {{Collective enhancement in dissipative quantum batteries}},
  author = {Carrasco, Javier and Maze, Jer\'onimo R. and Hermann-Avigliano, Carla and Barra, Felipe},
  journal = {Phys. Rev. E},
  volume = {105},
  issue = {6},
  pages = {064119},
  numpages = {6},
  year = {2022},
  month = {Jun},
  publisher = {American Physical Society},
  doi = {10.1103/PhysRevE.105.064119},
  url = {https://link.aps.org/doi/10.1103/PhysRevE.105.064119}
}

@article{Joshi22,
  title = {{Experimental investigation of a quantum battery using star-topology NMR spin systems}},
  author = {Joshi, Jitendra and Mahesh, T. S.},
  journal = {Phys. Rev. A},
  volume = {106},
  issue = {4},
  pages = {042601},
  numpages = {8},
  year = {2022},
  month = {Oct},
  publisher = {American Physical Society},
  doi = {10.1103/PhysRevA.106.042601},
  url = {https://link.aps.org/doi/10.1103/PhysRevA.106.042601}
}

@article{Cruz22,
doi = {10.1088/2058-9565/ac57f3},
url = {https://doi.org/10.1088/2058-9565/ac57f3},
year = {2022},
month = {mar},
publisher = {IOP Publishing},
volume = {7},
number = {2},
pages = {025020},
author = {Cruz, Clebson and Anka, Maron F and Reis, Mario S and Bachelard, Romain and Santos, Alan C},
title = {{Quantum battery based on quantum discord at room temperature}},
journal = {Quantum Science and Technology}
}

@article{Seah21,
  title = {{Quantum Speed-Up in Collisional Battery Charging}},
  author = {Seah, Stella and Perarnau-Llobet, Mart\'{\i} and Haack, G\'eraldine and Brunner, Nicolas and Nimmrichter, Stefan},
  journal = {Phys. Rev. Lett.},
  volume = {127},
  issue = {10},
  pages = {100601},
  numpages = {6},
  year = {2021},
  month = {Aug},
  publisher = {American Physical Society},
  doi = {10.1103/PhysRevLett.127.100601},
  url = {https://link.aps.org/doi/10.1103/PhysRevLett.127.100601}
}

@article{Shaghaghi22,
doi = {10.1088/2058-9565/ac8829},
url = {https://doi.org/10.1088/2058-9565/ac8829},
year = {2022},
month = {aug},
publisher = {IOP Publishing},
volume = {7},
number = {4},
pages = {04LT01},
author = {Shaghaghi, Vahid and Singh, Varinder and Benenti, Giuliano and Rosa, Dario},
title = {{Micromasers as quantum batteries}},
journal = {Quantum Science and Technology}
}

@Article{Shaghaghi23,
AUTHOR = {Shaghaghi, Vahid and Singh, Varinder and Carrega, Matteo and Rosa, Dario and Benenti, Giuliano},
title = {{Lossy Micromaser Battery: Almost Pure States in the Jaynes–Cummings Regime}},
JOURNAL = {Entropy},
VOLUME = {25},
YEAR = {2023},
NUMBER = {3},
ARTICLE-NUMBER = {430},
URL = {https://www.mdpi.com/1099-4300/25/3/430},
PubMedID = {36981319},
ISSN = {1099-4300},
DOI = {10.3390/e25030430}
}

@Article{Massa25,
AUTHOR = {Massa, Nicolò and Cavaliere, Fabio and Ferraro, Dario},
title = {{The Collisional Charging of a Transmon Quantum Battery}},
JOURNAL = {Batteries},
VOLUME = {11},
YEAR = {2025},
NUMBER = {7},
ARTICLE-NUMBER = {240},
URL = {https://www.mdpi.com/2313-0105/11/7/240},
ISSN = {2313-0105},
DOI = {10.3390/batteries11070240}
}

@article{Hu22,
doi = {10.1088/2058-9565/ac8444},
url = {https://doi.org/10.1088/2058-9565/ac8444},
year = {2022},
month = {aug},
publisher = {IOP Publishing},
volume = {7},
number = {4},
pages = {045018},
author = {Hu, Chang-Kang and Qiu, Jiawei and Souza, Paulo J P and Yuan, Jiahao and Zhou, Yuxuan and Zhang, Libo and Chu, Ji and Pan, Xianchuang and Hu, Ling and Li, Jian and Xu, Yuan and Zhong, Youpeng and Liu, Song and Yan, Fei and Tan, Dian and Bachelard, R and Villas-Boas, C J and Santos, Alan C and Yu, Dapeng},
title = {{Optimal charging of a superconducting quantum battery}},
journal = {Quantum Science and Technology}
}

@article{Gemme24,
  title = {{Qutrit quantum battery: Comparing different charging protocols}},
  author = {Gemme, Giulia and Grossi, Michele and Vallecorsa, Sofia and Sassetti, Maura and Ferraro, Dario},
  journal = {Phys. Rev. Res.},
  volume = {6},
  issue = {2},
  pages = {023091},
  numpages = {13},
  year = {2024},
  month = {Apr},
  publisher = {American Physical Society},
  doi = {10.1103/PhysRevResearch.6.023091},
  url = {https://link.aps.org/doi/10.1103/PhysRevResearch.6.023091}
}

@article{Li25,
  title = {{Stable and efficient charging of superconducting capacitively shunted flux quantum batteries}},
  author = {Li, Li and Zhao, Si-Lu and Shi, Yun-Hao and Chen, Bing-Jie and Ruan, Xinhui and Liang, Gui-Han and Yuan, Wei-Ping and Song, Jia-Cheng and Deng, Cheng-Lin and Liu, Yu and Li, Tian-Ming and Liu, Zheng-He and Guo, Xue-Yi and Song, Xiaohui and Xu, Kai and Fan, Heng and Xiang, Zhongcheng and Zheng, Dongning},
  journal = {Phys. Rev. Appl.},
  volume = {24},
  issue = {5},
  pages = {054033},
  numpages = {16},
  year = {2025},
  month = {Nov},
  publisher = {American Physical Society},
  doi = {10.1103/y3qx-cs3r},
  url = {https://link.aps.org/doi/10.1103/y3qx-cs3r}
}

@article{Morrone23,
doi = {10.1088/2058-9565/accca4},
url = {https://doi.org/10.1088/2058-9565/accca4},
year = {2023},
month = {may},
publisher = {IOP Publishing},
volume = {8},
number = {3},
pages = {035007},
author = {Morrone, Daniele and Rossi, Matteo A C and Smirne, Andrea and Genoni, Marco G},
title = {{Charging a quantum battery in a non-Markovian environment: a collisional model approach}},
journal = {Quantum Science and Technology}
}

@article{Elyasi25,
doi = {10.1088/2058-9565/adae2d},
url = {https://doi.org/10.1088/2058-9565/adae2d},
year = {2025},
month = {feb},
publisher = {IOP Publishing},
volume = {10},
number = {2},
pages = {025017},
author = {Navid Elyasi, Seyed and Rossi, Matteo A C and Genoni, Marco G},
title = {{Experimental simulation of daemonic work extraction in open quantum batteries on a digital quantum computer}},
journal = {Quantum Science and Technology}
}

@article{Xiang13,
  title = {{Hybrid quantum circuits: Superconducting circuits interacting with other quantum systems}},
  author = {Xiang, Ze-Liang and Ashhab, Sahel and You, J. Q. and Nori, Franco},
  journal = {Rev. Mod. Phys.},
  volume = {85},
  issue = {2},
  pages = {623--653},
  numpages = {0},
  year = {2013},
  month = {Apr},
  publisher = {American Physical Society},
  doi = {10.1103/RevModPhys.85.623},
  url = {https://link.aps.org/doi/10.1103/RevModPhys.85.623}
}

@article{Krantz19,
    author = {Krantz, P. and Kjaergaard, M. and Yan, F. and Orlando, T. P. and Gustavsson, S. and Oliver, W. D.},
    title = {{A quantum engineer's guide to superconducting qubits}},
    journal = {Applied Physics Reviews},
    volume = {6},
    number = {2},
    pages = {021318},
    year = {2019},
    month = {06},
    issn = {1931-9401},
    doi = {10.1063/1.5089550},
    url = {https://doi.org/10.1063/1.5089550},
}

@article{Lu25,
  title = {{Topological Quantum Batteries}},
  author = {Lu, Zhi-Guang and Tian, Guoqing and L\"u, Xin-You and Shang, Cheng},
  journal = {Phys. Rev. Lett.},
  volume = {134},
  issue = {18},
  pages = {180401},
  numpages = {8},
  year = {2025},
  month = {May},
  publisher = {American Physical Society},
  doi = {10.1103/PhysRevLett.134.180401},
  url = {https://link.aps.org/doi/10.1103/PhysRevLett.134.180401}
}

@Article{Wang26,
title = {{Research progress of quantum battery}},
journal = {Acta Physica Sinica},
volume = {75},
number = {4},
pages = {},
year = {2026},
issn = {1000-3290},
doi = {10.7498/aps.75.20251507},	
url = {https://wulixb.iphy.ac.cn/en/article/doi/10.7498/aps.75.20251507},
author = {WANG Lu and WU Fenglin and LI Nana and GUO Senyan and FAN Hao and LIU Shuqian and LIU Siyuan}
}

@misc{Sharma25,
      title = {{{Quadratic power enhancement in extended Dicke quantum battery}}}, 
      author={Harsh Sharma and Himadri Shekhar Dhar},
      year={2025},
      eprint={2512.15607},
      archivePrefix={arXiv},
      primaryClass={quant-ph},
      url={https://arxiv.org/abs/2512.15607}, 
}

@article{Ho26,
      title = {{Boosting the Performance of a Lipkin-Meshkov-Glick Quantum Battery via Symmetry-Breaking Quenches and Bosonic Baths}}, 
  author = {Ho, Le Bin and Hoang, Duc Tuan and Anh-Tai, Tran Duong and Busch, Thomas and Fogarty, Thom{\'a}s},
  journal = {Phys. Rev. A},
  volume = {114},
  issue = {2},
  pages = {022210},
  numpages = {15},
  year = {2026},
  month = {Aug},
  publisher = {American Physical Society},
  doi = {10.1103/8yq7-jp7r},
  url = {https://link.aps.org/doi/10.1103/8yq7-jp7r},
}

@misc{bonasera_arxiv_2026,
  title        = {{From Valley Filtering to Superconducting Diode Effect in Spin-Orbit Coupled Graphene Junctions}},
  author       = {F. Bonasera and G. A. Falci and E. Paladino and F. M. D. Pellegrino},
  year         = {2026},
  eprint       = {2607.06303},
  archivePrefix= {arXiv},
  primaryClass = {cond-mat.mes-hall},
  note         = {arXiv:2607.06303}
}

@article{carrega_njp_2020,
  author  = {Carrega, Matteo and Crescente, Alba and Ferraro, Dario and Sassetti, Maura},
  title   = {Dissipative dynamics of an open quantum battery},
  journal = {New Journal of Physics},
  volume  = {22},
  number  = {8},
  pages   = {083085},
  year    = {2020},
  doi     = {10.1088/1367-2630/abaa01}
}

@article{shastri_npj_2025,
  author    = {Rahul Shastri and Chao Jiang and Guo-Hua Xu and B. Prasanna Venkatesh and Gentaro Watanabe},
  title     = {Dephasing enabled fast charging of quantum batteries},
  journal   = {npj Quantum Information},
  volume    = {11},
  number    = {1},
  pages     = {9},
  year      = {2025},
  doi       = {10.1038/s41534-025-00959-5}
}

@misc{Varrica2026Dataset,
  author    = {Varrica, V. and
               Gemme, G. and
               Pellegrino, F. M. D. and
               Paladino, E. and
               Sassetti, M. and
               Ferraro, D.},
  note      = {{Quantum Batteries in two-dimensional material-based Josephson Junctions} [Dataset], DOI: 10.5281/zenodo.22639175}, 
  title     = {Quantum Batteries in two-dimensional material-based Josephson Junctions},
  year      = {2026},
  publisher = {Zenodo},
  type      = {Dataset},
  doi       = {10.5281/zenodo.22639175},
  url       = {https://doi.org/10.5281/zenodo.22639175}
}

\end{document}